\documentclass{sig-alternate-10pt}

\usepackage{times}
\usepackage{subfigure}
\usepackage[hyphens]{url}
\usepackage{epsfig}
\usepackage{graphics,xspace}
\usepackage{multirow} 
\usepackage{graphicx}
\usepackage{amsmath}

\newcommand{\Comment}[1]{{\bf #1}}
\renewcommand{\paragraph}[1]{\smallskip\noindent{\bf{#1.}}}

\begin{document}

\title{60 GHz Multi-Gigabit Indoor WLANs: Dream or Reality?}
%
%
%
%
%

\numberofauthors{3} 
%
\author{
%
%
\alignauthor
Swetank Kumar Saha\\
       \affaddr{University at Buffalo, SUNY}\\
       \email{swetankk@buffalo.edu}
\alignauthor
Viral Vijay Vira\\
       \affaddr{University at Buffalo, SUNY}\\
       \email{viralvij@buffalo.edu}
\alignauthor
Anuj Garg\\
       \affaddr{University at Buffalo, SUNY}\\
       \email{anujgarg@buffalo.edu}
\and  
\alignauthor
Dimitrios Koutsonikolas\\
       \affaddr{University at Buffalo, SUNY}\\
       \email{dimitrio@buffalo.edu}
}


\maketitle
\begin{abstract}

The millimeter-wave (mmWave) technology, recently standardized by IEEE
802.11ad, is emerging as an attractive alternative to the traditional
2.4/5GHz wireless systems, promising multi-Gigabit wireless
throughput. However, the high attenuation and vulnerability to
blockage of 60 GHz links have limited its applications (until
recently) to short-range, line-of-sight, static scenarios. On the
other hand, the question of whether it is feasible to build
general-purpose WLANs out of mmWave radios in dynamic indoor
environments with non-line-of-sight links remains largely unanswered.

In this paper, through extensive measurements with COTS 802.11ad
hardware in an indoor office environment, we investigate the question
of whether the mmWave technology, in spite of its unique propagation
characteristics, can serve as a viable choice for providing
multi-Gigabit ubiquitous wireless indoor connectivity. We study the
range of 60 GHz transmissions in indoor environments, the impact of
antenna height, location, orientation, and distance on 60 GHz
performance, the interaction among metrics from different layers of
the network stack, the increased opportunities for spatial reuse, and
the impact of human blockage. Our results reveal a number of
challenges that we have to address for 60 GHz multi-gigabit WLANs to
become a reality.

\end{abstract}

\section{Introduction}
\label{sec:intro}


We experience today an explosion in wireless network traffic driven by
the rapidly growing number of mobile devices and bandwidth hungry
applications. Industry research predicts that the aggregate bandwidth
demands will increase by 1000x by 2020~\cite{infographic}. Although
intelligent spectrum adaptation in WiFi and cellular networks can
offer short-term solutions, it is unlikely to completely solve the
problem of high bandwidth demands in the longer term. The scalability
of any solution in the lower frequency bands (LTE, WiFi, TV white
spaces) is physically limited by form factor constraints, which put a
hard limit to the number of antennas that can be accommodated on an AP
and/or a client device, and call for alternative approaches.

One alternative approach which has recently attracted significant
interest in the industry is the use of millimeter-wave (mmWave) radios
in the 60 GHz band, which provides 7 GHz of {\em unlicensed} spectrum,
and is supported by the new IEEE 802.11ad
standard~\cite{802_11ad_std}. 802.11ad defines three 2.16 GHz channels
and offers bitrates between 385 Mbps and 6.76 Gbps. Further, the
order-of-magnitude shorter wavelengths in the 60 GHz band compared to
the 2.4/5 GHz bands makes it possible to pack a very large number of
antenna arrays into small factors, enabling highly directional
transmissions. Narrow directional beams limit interference and allow
for very dense deployments, significantly improving spatial reuse.

The caveat is the high attenuation and vulnerability to
blockage of 60 GHz links. Since free-space loss scales up with the
square of the carrier frequency, the propagation loss at 60 GHz is
21.6 dB worse than at 5 GHz. Furthermore, due to their small
wavelength, signals in the 60 GHz band are easily blocked by obstacles
such as walls, furniture, or humans. For example, a human in the
Line-of-Sight (LOS) between the transmitter and the receiver can
attenuate the signal by 20-30 dB~\cite{singh:jsac2009}, resulting in
link outage. While the use of electronically steerable
antenna arrays can theoretically overcome link outages, the overhead
of re-beamforming may counter-balance the potential
gains~\cite{tie:pam2012,zhu:mobicom2014,sur:sigmetrics2015}. Moreover,
directional transmissions pose a great challenge to mobility as mobile
clients and APs may have to continuously re-arrange their beams.

Due to these reasons, the use of the 60 GHz technology has been
limited (until recently) to short-range, (mostly) LOS, static
scenarios, e.g., for HD video streaming in wireless personal area
networks (WPANs)~\cite{wirelessHD}, wireless
docking~\cite{d5000}, or for augmenting data center networks with high
capacity wireless
links~\cite{halperin:sigcomm2011,zhou:sigcomm2012,shin:ton2013,zhu:mobicom2014}. These
scenarios are characterized by open spaces providing direct LOS paths
but largely devoid of reflective and/or obstructive surfaces and
objects. The absence of phenomena such as reflection or multipath
makes such environments easy to model as they exhibit near-free space
propagation properties. On the other hand, {\em the question of
  whether it is feasible to build general-purpose WLANs out of mmWave
  radios in environments with NLOS links and in the presence of user
  mobility remains largely unanswered}.

In this paper, we investigate the feasibility of building {\em
  general-purpose indoor 60 GHz WLANs for enterprise environments,
  that will offer always-on connectivity but an order of magnitude
  higher throughput than today's 2.4/5 GHz WLANs}. Compared to WPAN or
datacenter environments, the typical indoor enterprise WLAN
environment is highly complex, with many objects/surfaces that can
attenuate, completely block, or reflect the signal, making it harder
to predict link behavior. Our main contributions and findings are
summarized as follows:

\noindent(1) This paper is one of the first to investigate the
feasibility of building general-purpose, indoor, 60 GHz WLANs by
conducting an extensive measurement study in a typical academic office
building. {\em In contrast to previous works which measured
  performance at a single layer (typically PHY) using custom,
  non-standard compliant hardware, we investigate performance across
  multiple layers of the protocol stack using 802.11ad-compliant COTS
  devices}. Although the use of commodity hardware limits our access
to some PHY layer information (e.g., RSS) and our control on certain
protocol features (e.g., rate adaptation, beamforming), it allows us
to get a much closer idea of how the 802.11ad technology would behave
when used to build full-scale WLANs. Our results differ significantly
in several aspects from both theoretical modeling and experimental
results reported in the past, suggesting that link characterization
based on non-standard-compliant hardware or simple theoretical models
cannot always provide deep insights in a complex environment.

\noindent(2) We study the performance of 60 GHz links at multiple
locations and orientations (Section~\ref{sec:loc_or}). Our results
confirm that high-throughput 60GHz communication is feasible at
various setups typical of an indoor WLAN environment (corridors,
halls, labs, through walls or glass). Moreover, communication is
possible in the case of antenna array misalignment, either via
beamsteering or through an NLOS link via reflection, but the
performance can vary significantly depending on the location and Tx-Rx
distance and can be much lower than in the case of optimal
orientation.

\noindent(3) We study the impact of Tx-Rx distance on link quality
indicators -- RSSI, PHY data rate -- and higher layer performance --
TCP throughput -- (Section~\ref{sec:distance}). We find communication
ranges much longer than those reported by previous studies with the
same hardware. However, performance is unpredictable and highly
dependent on the environment (type and number of reflective
surfaces). In particular, we find that signal propagation in indoor
WLAN environments cannot be characterized by simple log distance path
loss models, which have been extensively used in 802.11ad
simulators~\cite{halperin:sigcomm2011,zhou:sigcomm2012,tie:pam2012,zhu2:mobicom2014}.

\noindent(4) We investigate the relationship among RSSI, PHY data
rate, and TCP throughput (Section~\ref{sec:predict}). We find that
RSSI can only serve as a weak indicator of PHY data rate and TCP
throughput and only at certain locations, but not across
locations. Hence, translating signal strength to PHY data rate, a
common practice in recent measurement
studies~\cite{zhu:mobicom2014,sur:sigmetrics2015}, can yield
inaccurate results in typical indoor WLAN environments. Further, we
find that PHY data rate is not always a good indicator of higher layer
performance.

\noindent(5) We examine the benefits brought by directional 60 GHz
links in terms of spatial reuse (Section~\ref{sec:reuse}). Our results
reveal that, even with imperfect hardware -- wide beams and sidelobes
-- 802.11ad allows for much higher spatial reuse compared to 802.11ac.

\noindent(6) We study the impact of human blockage on the performance
of 60 GHz links (Section~\ref{sec:human}). Our results show that human
blockage, especially by humans standing near the Tx or Rx, presents a
major challenge for 60 GHz links in indoor environments. However, a
client served by two APs simultaneously can maintain 100\% uptime and
high throughput.

Overall, our results show that the 60 GHz technology can be a viable
option for providing ubiquitous indoor multi-gigabit wireless
connectivity, but also uncover a number of challenges that need to be
addressed for this vision to become a reality.

\section{Related Work}
\label{sec:related}

\noindent{\bf 60 GHz in indoor environments} Our work is not the first
to investigate the feasibility of 60GHz technology in indoor
WLANs. Initial experimental work focused on measuring and modeling
channel propagation characteristics using dedicated channel sounding
hardware
(e.g.,~\cite{smulders:ecej1997,xu:jsac2002,anderson:twc2004,smulders:tap2009,
  maktsev:jsac2009}). A few works also
measured~\cite{collonge:twc2004}, simulated~\cite{sato:vtc1998}, or
studied analytically~\cite{dong:letters2012} the impact of human
blockage in indoor environments. Tie et al.~\cite{tie:pam2012} studied
link level performance of 60GHz links with respect to blockage and
antenna orientation. However, they used custom designed non-802.11ad
hardware and measured performance of IP-over-wireless-HDMI. In
contrast to these works, we are using COTS 802.11ad-compliant hardware
equipped with phased antenna arrays and measure performance via data
transfers over real transport layer protocols (TCP/UDP).

More recently, Sur et al.~\cite{sur:sigmetrics2015} conducted a
link-level profiling of indoor 60 GHz links, using a custom
software-radio platform (WiMi). Their study offers many valuable
insights, in particular about the potential capabilities and
limitations of flexible beams. However, WiMi uses a small channel
width of only 245 MHz and thus, it cannot achieve Gbps data
rates. Hence, findings in this work are extrapolated from RSS
measurements in narrow channels and they may not reflect the behavior
of real 802.11ad links. Moreover, throughput values are not obtained
through real data transfers but are translated from RSS and noise
floor measurements in narrow channels using an 802.11ad specific rate
table. In contrast, in our study we directly measure performance of
COTS 802.11ad devices.

More importantly, {\em the main limitation of all the above studies is
  that they target a single layer of the protocol stack (PHY, link, or
  transport)}. In contrast, our goals in this paper are to (i) measure
performance {\em across different layers} using COTS 802.11ad hardware
and (ii) understand the {\em interactions} between different
performance metrics.

Recent work also has argued for the use of 60GHz technology to augment
datacenters~\cite{halperin:sigcomm2011,zhou:sigcomm2012,zhu:mobicom2014}
and demonstrated the feasibility of this approach using both expensive
proprietary
devices~\cite{halperin:sigcomm2011,zhou:sigcomm2012,zhu:mobicom2014}
and the same cheap off-the-shelf hardware we use in this
paper~\cite{zhu2:mobicom2014}. The datacenter environment, with static
LOS links established on top of TOR switches, is very different from
the complex indoor enterprise WLAN environment, and several of our
findings are very different from the findings of these works.

\noindent{\bf 60 GHz in outdoor environments} Channel sounding
measurements have also been conducted in outdoor environments
\cite{langen:pimrc1994,smulders:ecej1997,rappaport:tap2013}. Similarly,
a few works simulated the impact of human blockage in such
environments~\cite{abouelseoud:vtc2013}. Several works have considered
the use of 60 GHz technology for building outdoor mesh networks for
backhaul, e.g., in 5G
networks~\cite{mudumbai:infocom-mini2009,signh:infocom2010,singh:ton2011,taori:commagazine2015};
all these proposals have only been evaluated in simulations. Finally,
recent work experimentally demonstrated the feasibility of 60
GHz-based outdoor picocells~\cite{zhu:mobicom2014} using both the
hardware we use in this work and proprietary non-802.11ad
hardware. Outdoor environments differ significantly from the one we
are concerned with, as also pointed out in~\cite{sur:sigmetrics2015},
and several observations reported in these works do not hold for our
use-case.

\section{Measurement methodology}

\subsection{Hardware}
Our 802.11ad link setup consists of two COTS devices: a Dell Latitude
E420 laptop equipped with a Wilocity wil6210 802.11ad
radio~\cite{wil_radio} and a Dell Wireless Dock
D5000~\cite{d5000}. The dock has an 802.11ad wireless interface and
acts as an AP. Another laptop is connected to the dock through a
Gigabit Ethernet interface to generate/receive TCP traffic. The use of
the Ethernet interface limits the throughput in our experiments to 1
Gbps, even though the wireless link itself is capable of much higher
speeds. Hence, we also use the dominant PHY data rate as an indicator
of the maximum achievable throughput. The Wilocity radios do not allow
us to control the PHY layer data rate and use their own rate
adaptation algorithm\footnote{The following PHY data rates are
  supported by our cards (in Mbps): 385, 770, 1155, 1540, 1925, 2310,
  3080, 3850.}. Beam properties (beam direction and width) are
controlled by an in-built beamforming mechanism. Further, in case the
link between the dock and laptop is blocked/broken, the dock
automatically searches for an alternative link (e.g., through
reflection) to re-establish the connection. The Wilocity radios export
to the user-space the current PHY data rate and and an RSSI value
between 0 and 100 indicating current link quality.

Although IEEE 802.11ad specifies antenna beams as narrow as $2.86^o$,
which can be achieved, for example, with a 10x10 antenna
array~\cite{zhu:mobicom2014}, the Wilocity radios use 2x8 antenna
arrays with a main beamwidth of
$30^o-40^o$~\cite{zhu2:mobicom2014,sur:sigmetrics2015}. Hence, our
throughput and range measurements are probably lower bounds of the
achievable performance of 802.11ad links. However, while narrow beam
antennas can greatly extend range, recent
work~\cite{sur:sigmetrics2015} has shown that they yield poor
performance with client mobility and human blockage. Hence, we believe
that future WLANs may use wider beams.

\subsection{Locations}
For most of our experiments, we chose two kinds of environments/
locations: \textit{Hall} and \textit{Corridor}, to capture the diverse scenarios that
are likely to occur in a office environment. Both locations are inside
an academic building (Davis Hall) at the University at Buffalo, SUNY.
The building houses many faculty offices, classrooms, labs, and open
student areas, representative of a dense office environment full of
furniture and other objects/surfaces that can attenuate, completely
block, or reflect the signal.

The first location is an open \textit{Hall} thinly populated by some
desks and chairs and a staircase from the floor above. This location
offers better conditions for emulating free-space propagation and LOS
links, reducing multipath effects and the chances of communication
through NLOS links via reflection. The ceiling height is rather high
and thus it does not serve as an viable reflector. The second location
is a rather narrow \textit{Corridor} (only 1ft wide). This environment
offers ample opportunities for reflection/multipath from the walls on
the side, in addition to the floor and the ceiling.

Apart from these two locations, we also studied 60 GHz link behavior
through different commonly found materials (see Table \ref{tab:loc}) and some
typical setups inside the building with
different orientations of transmitter (Tx) and receiver (Rx) in
Section~\ref{sec:loc_or}. In addition, we performed control
experiments in an outdoor open environment, devoid of objects, to
better understand and explain some phenomena which we believe are
characteristic of an indoor environment.


\subsection{Methodology}
We used iperf3 to generate TCP traffic. Each experiment consists of a
10-second TCP and/or UDP session. All the results are the average of 5
sessions. All experiments were performed late night to remove the
possibility of human blockage. We study the effects of human blockage
on performance separately in Section~\ref{sec:human}. Unless stated
otherwise, the experiment environment consists of only static objects
present in the building environment.

\section{Transmitter Height}
\label{sec:height}

We begin by investigating the effect of dock/transmitter (Tx) height
on performance as measured at the transport layer. We pick two
different distances (8ft and 16ft) at each of the two locations and
measure TCP and UDP throughput with different Tx height. To decide
upon the optimal height for the Tx, we fix the height of the receiver
(Rx) to 2'6'' and vary the Tx height from 2'6'' to 6'6'' at the
interval of 1'.

\subsection{Rx Orientation}
In a real deployed WLAN, the AP's location and orientation will be
fixed. However, mobile clients can be assumed to have any possible
orientation with respect to the AP. We will discuss in detail in
Section~\ref{sec:or} how the relative orientation of the Tx and Rx
antenna array is an important factor in determining the link
performance. For this experiment, we chose four representative
orientations: 0, 4, 8 and 12 (see Table~\ref{tab:or}) for the Rx
and repeat our experiments at each one of them. Orientation \#0 is the
optimal case when both Tx and Rx antenna array are
aligned. Orientations \#4, 8 and 12 are obtained by rotating the Rx by
$90^{\circ}$, $180^{\circ}$ and $270^{\circ}$, respectively, while the
AP/dock orientation remains fixed.

\subsection{Results}

\begin{figure*}[]
\centering 
\subfigure[Hall (8 ft): TCP Throughput] {
\label{fig:height_f1_8-6_tcp}
\includegraphics[width=2.2in,height=1.6in,angle=0]{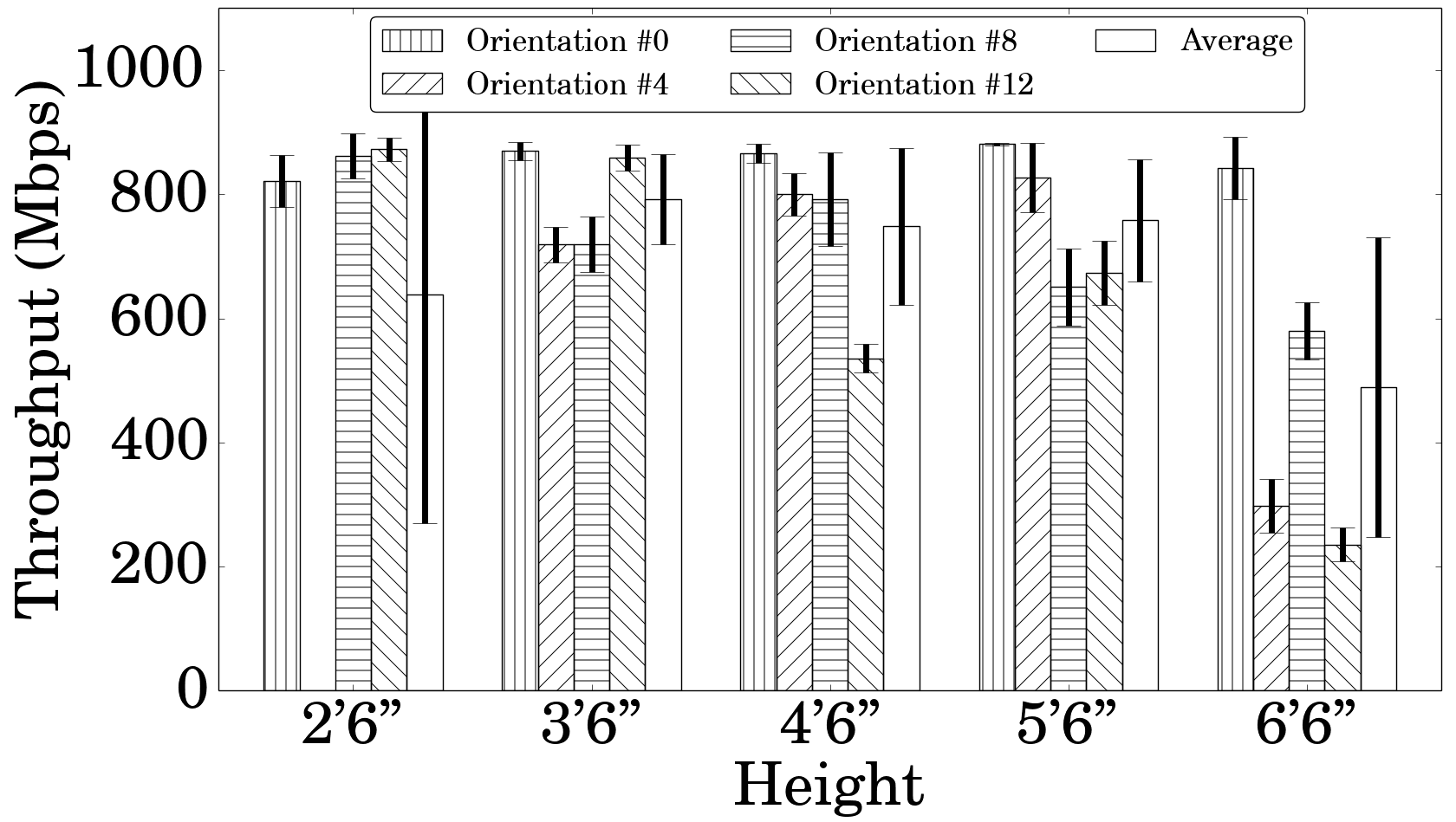} 
}
\subfigure[Corridor (8 ft): TCP Throughput] {
\label{fig:height_f3_8-6_tcp}
\includegraphics[width=2.2in,height=1.6in,angle=0]{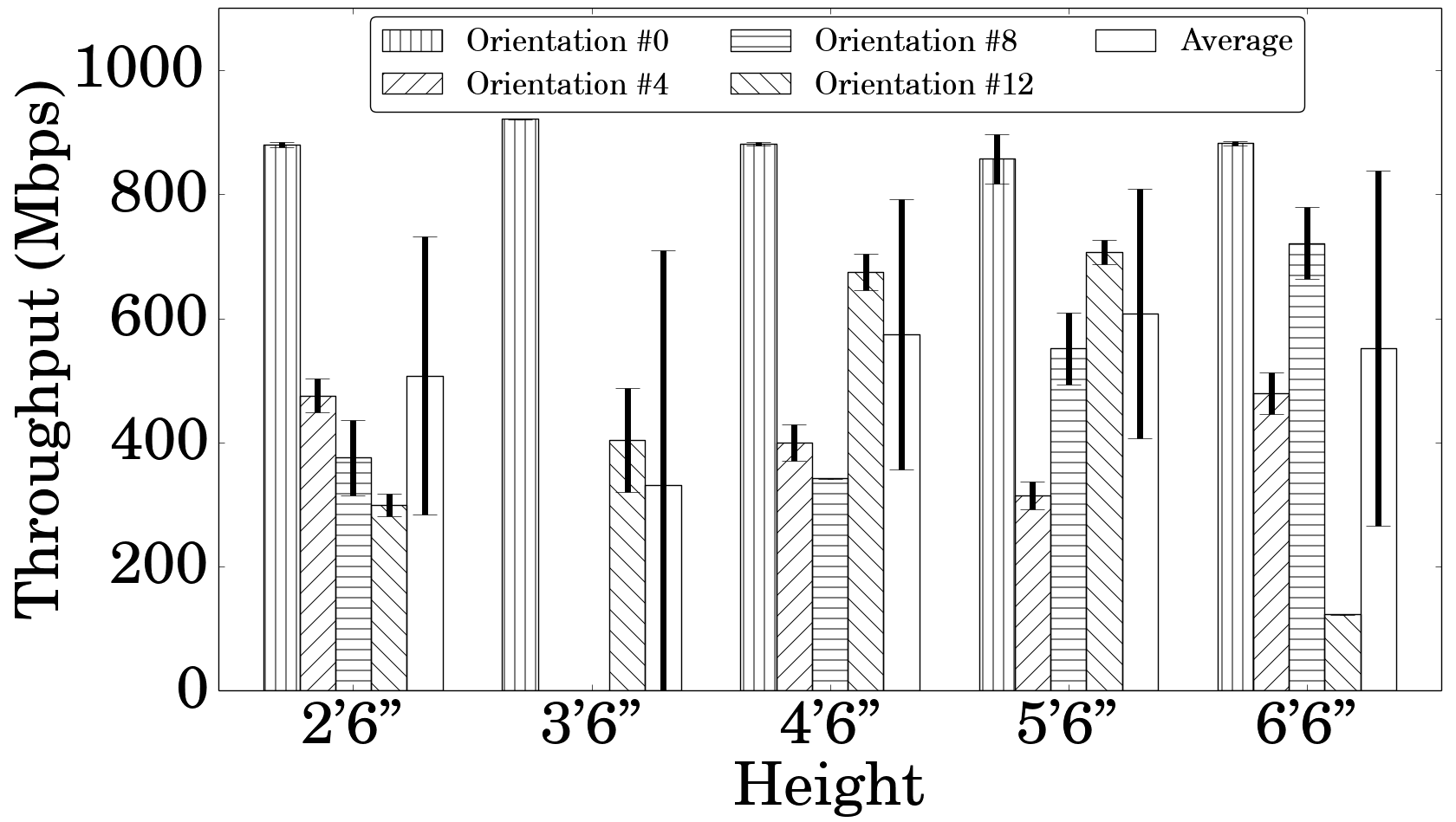}
} 
\subfigure[Hall (16 ft): TCP Throughput] {
\label{fig:height_f1_16-6_tcp}
\includegraphics[width=2.2in,height=1.6in,angle=0]{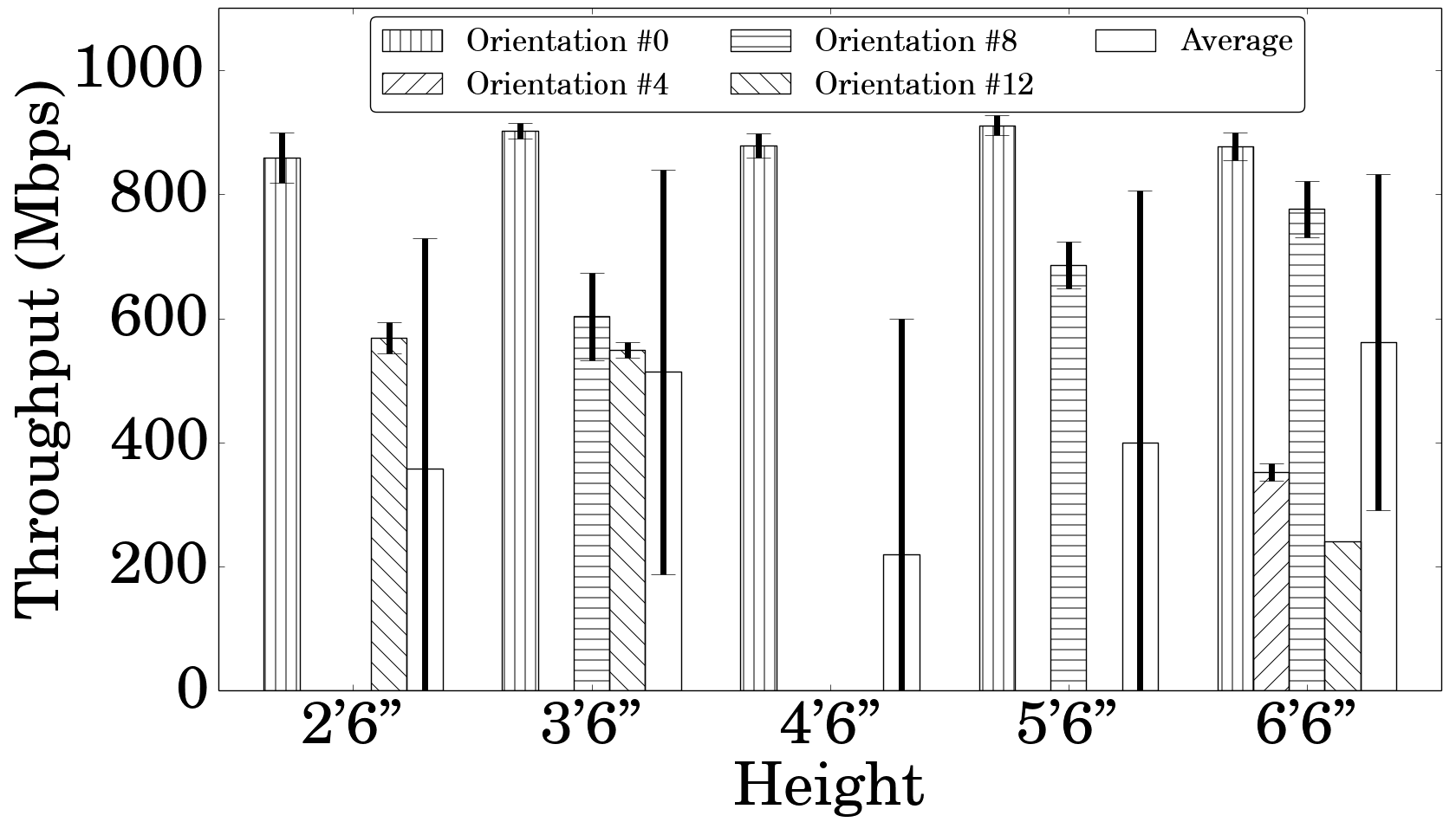} 
}

\subfigure[Hall (8 ft): UDP Throughput] {
\label{fig:height_f1_8-6_udp}
\includegraphics[width=2.2in,height=1.6in,angle=0]{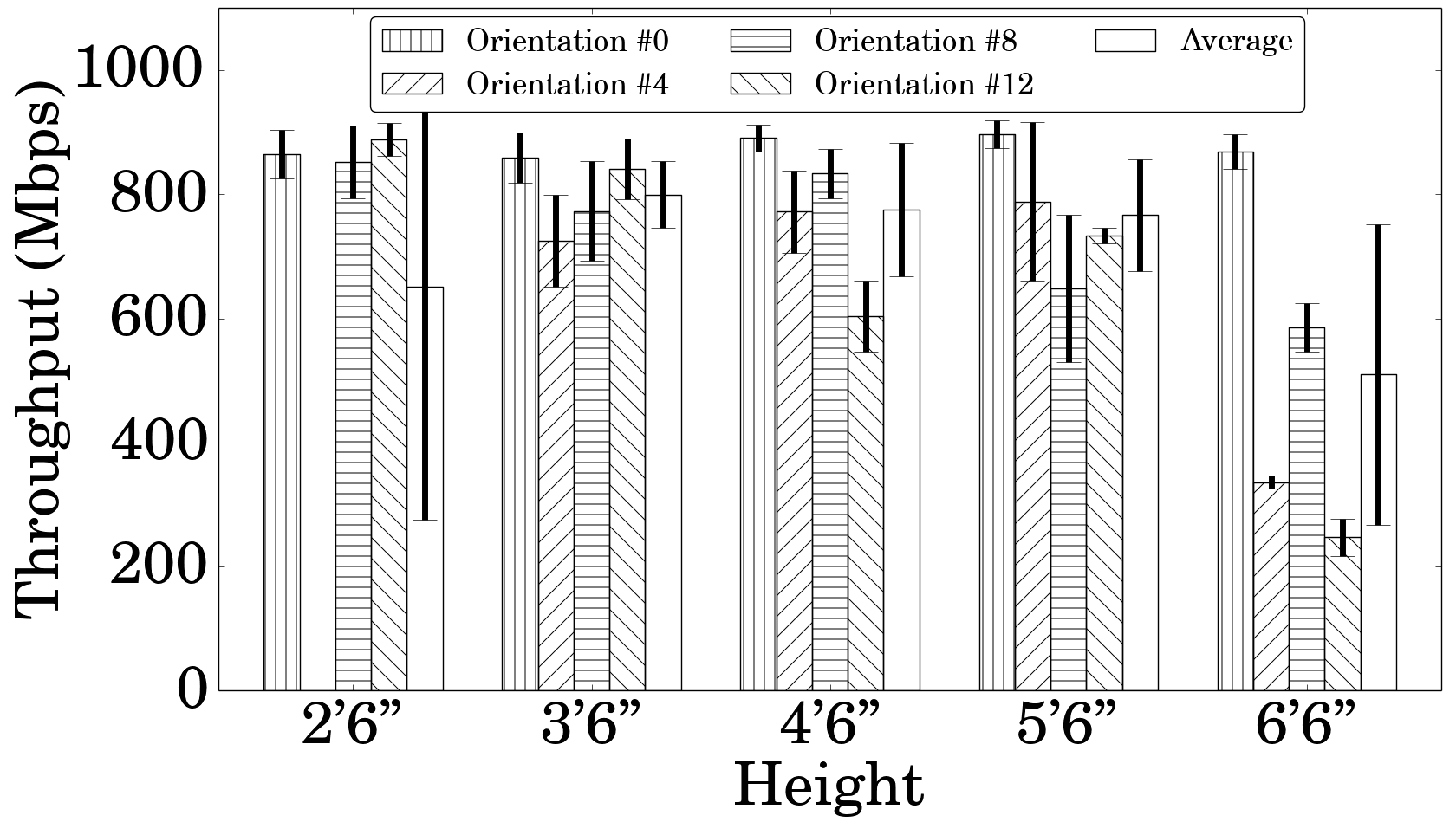} 
}
\subfigure[Corridor (8 ft): UDP Throughput] {
\label{fig:height_f3_8-6_udp}
\includegraphics[width=2.2in,height=1.6in,angle=0]{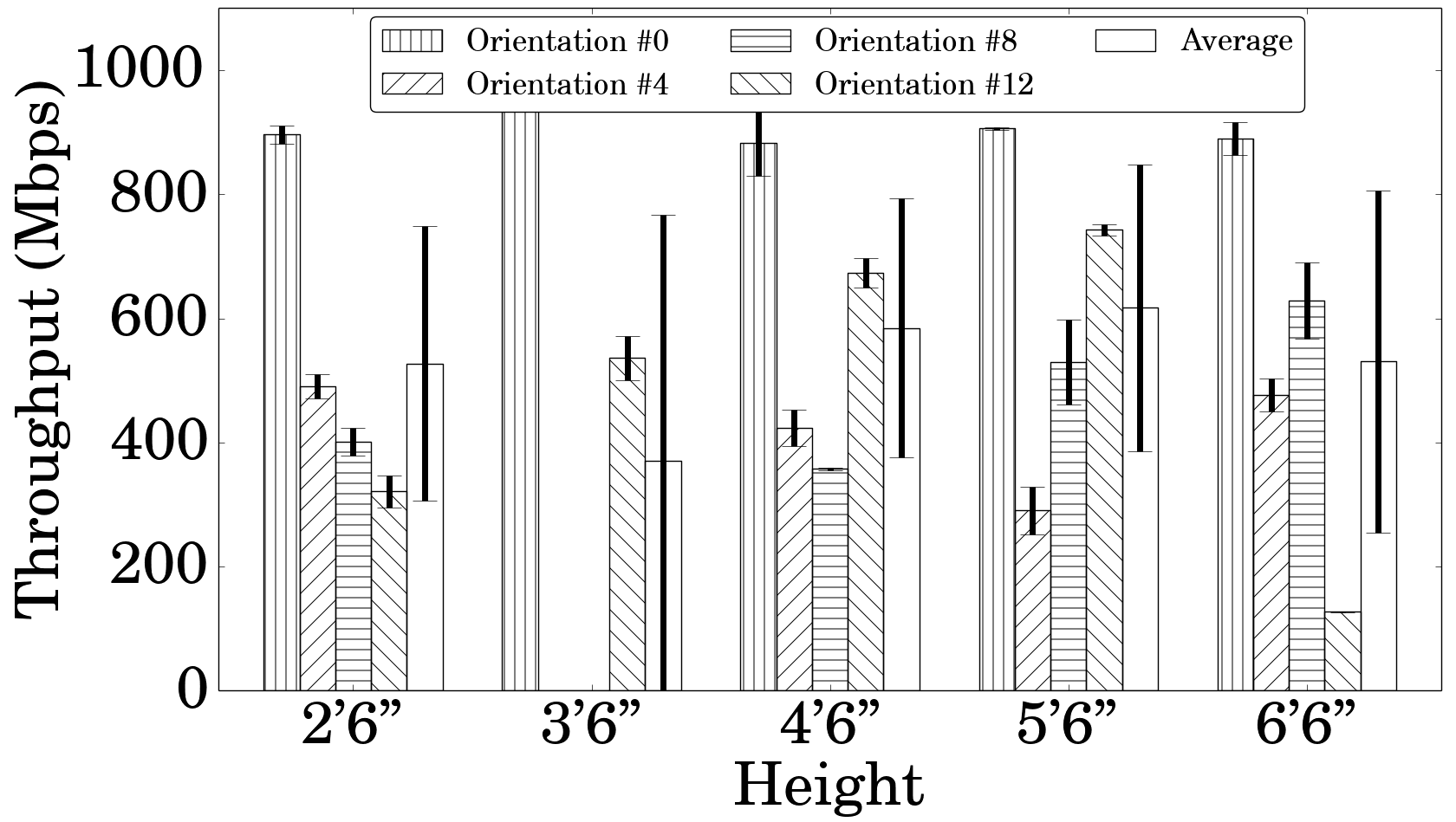}
} 
\subfigure[Corridor (16 ft): TCP Throughput] {
\label{fig:height_f3_16-6_tcp}
\includegraphics[width=2.2in,height=1.6in,angle=0]{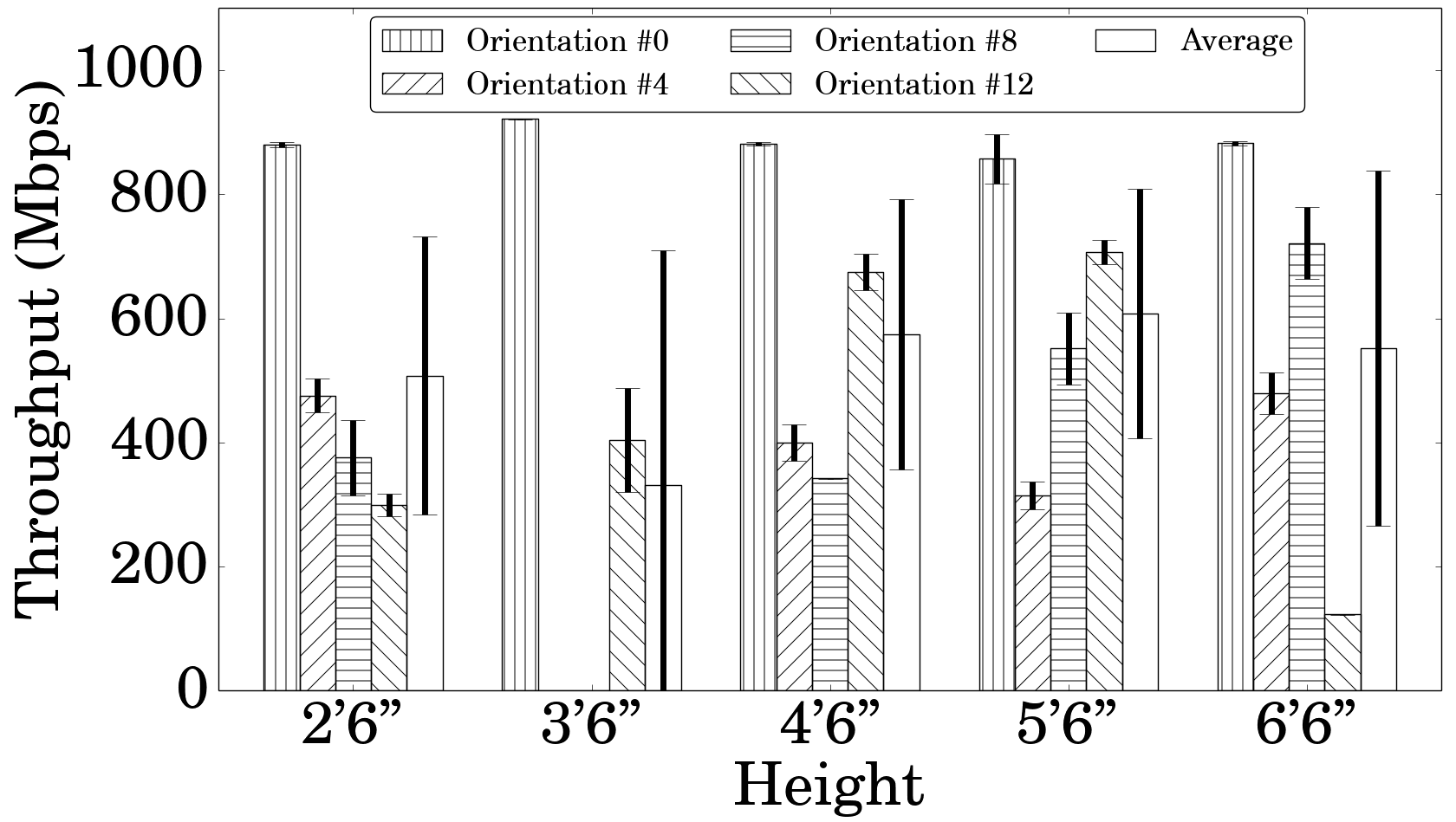} 
}
\vspace{-0.1in}
\caption{Variation of TCP/UDP throughput with Tx height.}

\label{fig:heights}
\end{figure*}

Figures \ref{fig:height_f1_8-6_tcp} and \ref{fig:height_f1_8-6_udp}
plot the TCP and UDP throughput achieved in the Hall for each of the
four orientations and an average across all orientations for different
Tx height. When both Tx and Rx are at the same height (2'6''),
orientation \#4 gives zero throughput, resulting in low average with a
large standard deviation. As we increase the height of Tx to 3'6'',
4'6'' and 5'6'', the average for all orientations approaches 800 Mbps,
indicating that larger Tx height is more favorable to link
performance. When the height is increased further to 6'6'', the
throughput for all orientations deteriorates significantly, giving an
average even lower than 2'6''. However, in contrast to 2'6'' Tx
height, note that 6'6'' Tx height can support all four orientations.

Figures \ref{fig:height_f3_8-6_tcp} and \ref{fig:height_f3_8-6_udp}
plot transport layer throughput for experiments done in the
Corridor. Here, the average throughput increases for heights 4'6'' and
5'6'' but it reduces slightly for 6'6''. The 3'6'' height, which
performed similarly to 4'6'' and 5'6'' in the hall, fails to provide
connectivity at orientations \#4 and \#8. In this case, 5'6'' seems to
be optimal height providing the highest average throughput.

To further ensure that we choose the best possible Tx height,
experiments at both locations were repeated with an increased distance
of 16' between the Tx and Rx. Figures \ref{fig:height_f1_16-6_tcp} and
\ref{fig:height_f3_16-6_tcp} plot the TCP throughput for Hall and
Corridor, respectively (we omit UDP results which are very similar to
TCP in the interest of space). In the Hall, 3'6'' provides a
throughput 100 Mbps higher than 5'6'' but in the Corridor, 5'6''
outperforms 3'6'' by more than 200 Mbps. Lastly, in the Corridor
environment, 5'6'' can support all four orientations while 3'6'' can
give non-zero throughput only for two orientations.

A general observation from these experiments is that {\em performance
  over 60 GHz indoor links is heavily affected by a number of factors
  -- location, orientation, Tx height, Tx-Rx distance -- and it is
  highly unpredictable; e.g., changing the Tx height by just 1' can
  improve or deteriorate TCP/UDP throughput by several hundreds of
  Mbps}. Hence, it is very hard to pick an optimal height for all
possible locations, antenna orientations, and link distances. Overall,
our experiments suggest that 5'6'' performs better in general,
although in some cases 3'6'' shows better performance. We decided in
favor of 5'6'' since previous studies have concluded that larger Tx
height results in propagation characteristics closer to the Friis
model~\cite{sur:sigmetrics2015} as it reduces the chances of multipath
caused by reflections from the floor. Also, from a perspective of
deployment of WLANs, a larger height for the AP is desirable for a
larger coverage distance/area. For the rest of the experiments, unless
stated otherwise, we use 5'6'' as the height of the transmitter.
 
\section{Performance across locations and orientations}
\label{sec:loc_or}
\vspace{-0.2in}
\begin{table}[h]
\centering
{\scriptsize
\caption{Measurement Locations}
\label{tab:loc}
\begin{tabular}{|c||c|c|}
\hline
\textbf{Location\#} & \textbf{Distance} & \textbf{Description} \\ \hline
0              & 8'6"                & Hall                   \\ \hline
1              & 16'                 & Hall                   \\ \hline
2              & 8'6"                & Corridor/Sym.          \\ \hline
3              & 8'6"                & Corridor/Asym.         \\ \hline
4              & 16'                 & Corridor/Asym.         \\ \hline
5              & 8'6"                & Wall                   \\ \hline
6              & 8'6"                & Glass                  \\ \hline
7              & 8'6"                & Corner                 \\ \hline
8              & 8'6"                & Lab                    \\ \hline
9              & 24'                 & Lab                    \\ \hline
\end{tabular}}
\end{table}
\vspace{-0.3in}
\begin{table}[h]
\centering
{\scriptsize
\caption{Measurement Orientations}
\label{tab:or}
\begin{tabular}{|c||c|c|c||c|c|}
\hline
\textbf{Orientation\#} & \textbf{Rx}            & \textbf{Tx}            & \textbf{Orientation\#} & \textbf{Rx}           & \textbf{Tx           } \\ \hline
0     & $\rightarrow$ & $\leftarrow$  & 8     & $\leftarrow$ & $\leftarrow$  \\ \hline
1     & $\rightarrow$ & $\downarrow$  & 9     & $\leftarrow$ & $\downarrow$  \\ \hline
2     & $\rightarrow$ & $\rightarrow$ & 10    & $\leftarrow$ & $\rightarrow$ \\ \hline
3     & $\rightarrow$ & $\uparrow$    & 11    & $\leftarrow$ & $\uparrow$    \\ \hline
4     & $\uparrow$    & $\leftarrow$  & 12    & $\downarrow$ & $\leftarrow$  \\ \hline
5     & $\uparrow$    & $\downarrow$  & 13    & $\downarrow$ & $\downarrow$  \\ \hline
6     & $\uparrow$    & $\rightarrow$ & 14    & $\downarrow$ & $\rightarrow$ \\ \hline
7     & $\uparrow$    & $\uparrow$    & 15    & $\downarrow$ & $\uparrow$    \\ \hline
\end{tabular}}
\end{table}

\vspace{-0.1in}
In this section, we study the impact of location and antenna array
orientation on the performance of 60 GHz links. In addition to the
Hall and Corridor locations, we study the performance in a number of
other setups inside the building, representative of typical scenarios
in an office environment, e.g., through different commonly found
materials (e.g. wood, glass) or around corners, with different Tx/Rx
orientations. We followed a methodology very similar to that
in~\cite{tie:pam2012} and performed multiple experiments at each of
the locations with same 16 different Tx/Rx orientations as
in~\cite{tie:pam2012} (see Table~\ref{tab:or} and Figure 4(b)
in~\cite{tie:pam2012}). There is at least one setup for each location
where the transmitter and receiver are placed rather close to each
other (around 8 ft). This ensures that we measure the best case
performance, when the signal does not attenuate severely as a function
of distance. At a few locations, we also present results for longer
distances for comparison.

\subsection{Impact of location}
\label{sec:loc}

\begin{figure*}[ht]
\centering
  \subfigure[RSSI.] {
  \label{fig:loc_rssi}
  \includegraphics[width=2.2in,height=1.6in,]{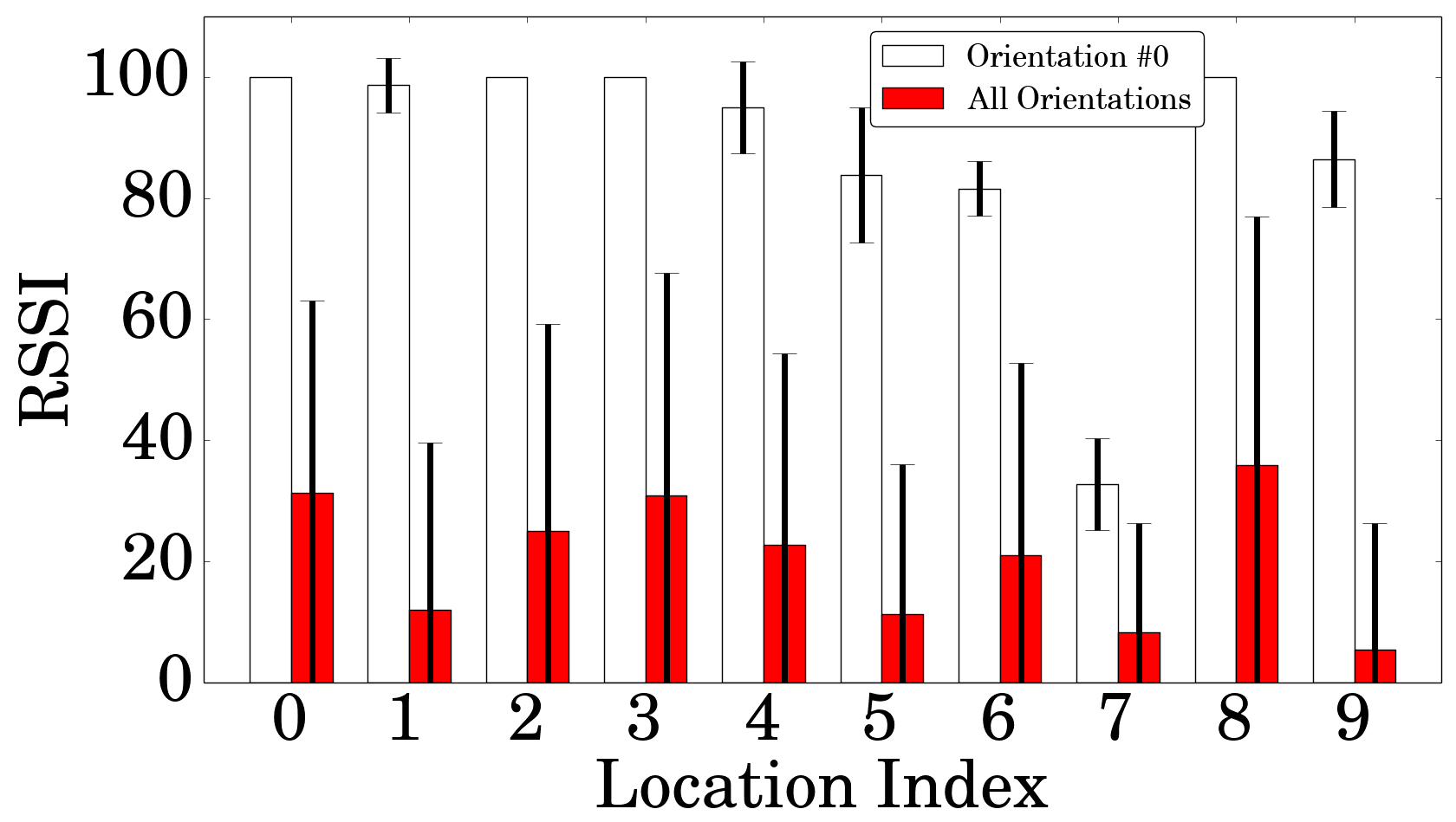} 
}
   \hspace{0.01in}
\subfigure[PHY data rate distribution.] {
  \label{fig:loc_rate}
  \includegraphics[width=2.2in,height=1.6in]{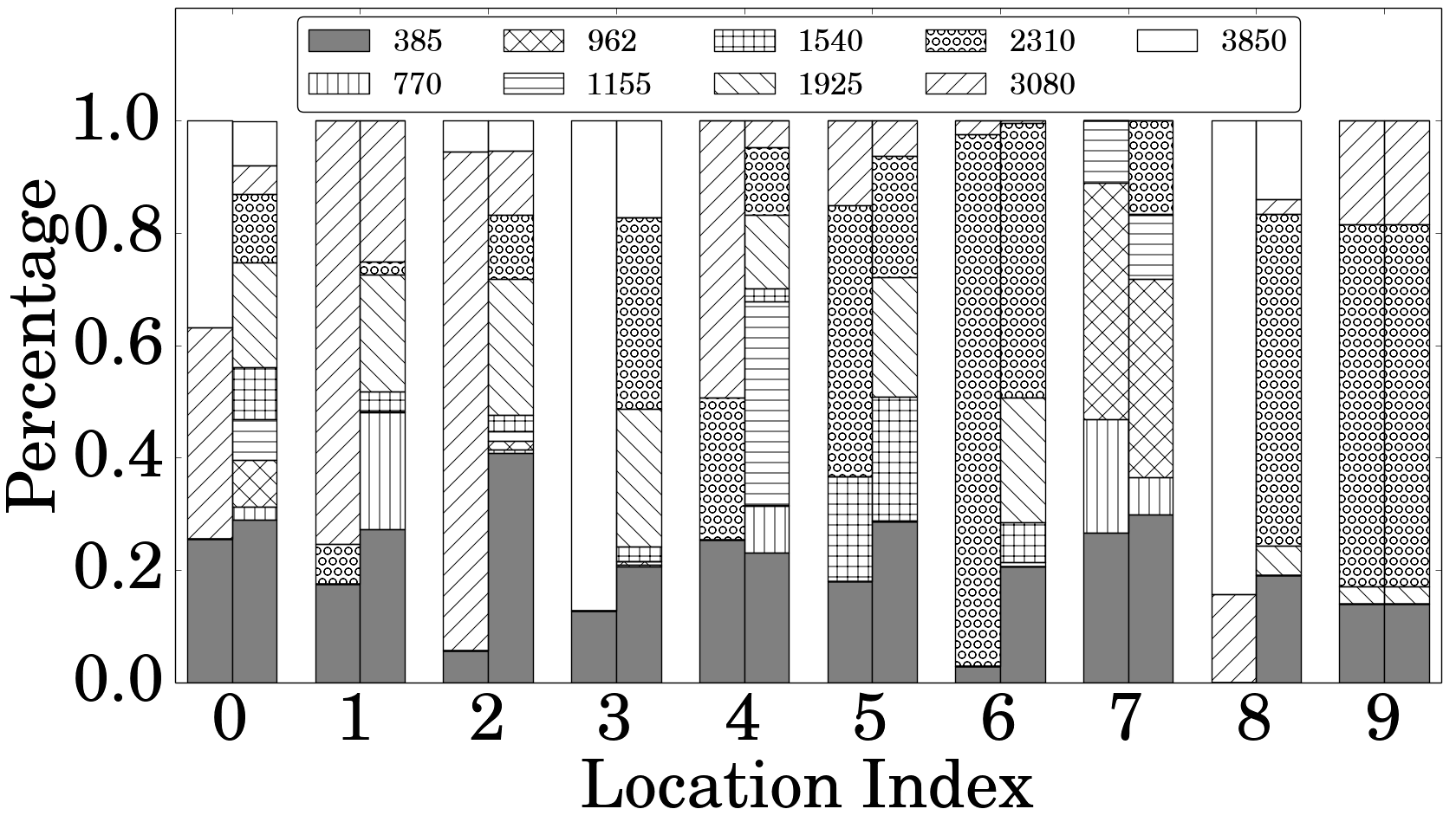}
} 
 \hspace{0.01in}
\subfigure[TCP throughput.] {
  \label{fig:loc_xput}
  \includegraphics[width=2.2in,height=1.6in]{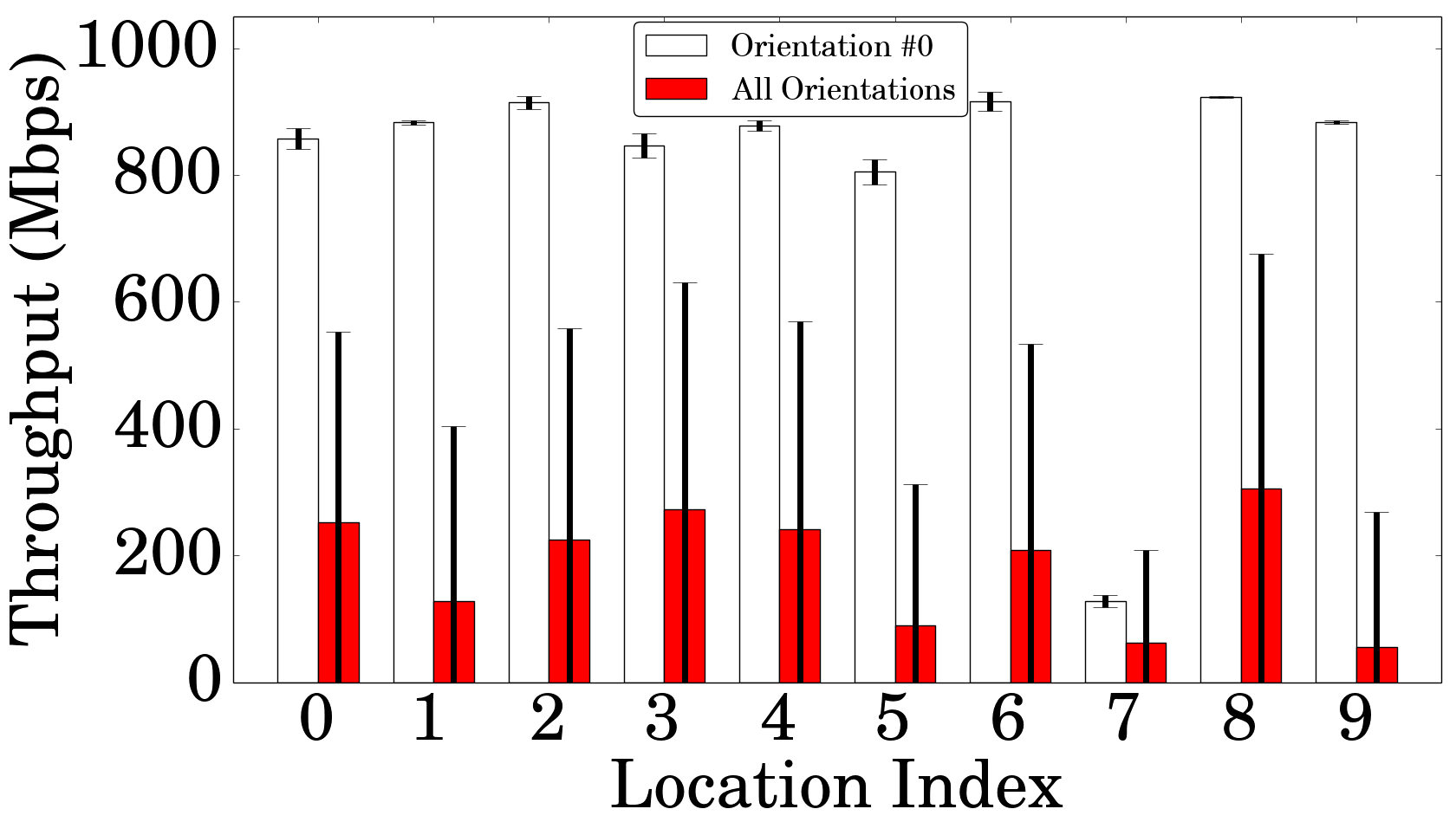}
 }
\vspace{-0.15in}
\caption{Performance across different locations.}
\vspace{-0.15in}
\label{fig:loc}
\end{figure*}

Figures~\ref{fig:loc_rssi},~\ref{fig:loc_rate},~\ref{fig:loc_xput}
plot the average RSSI, the selected PHY rates, and the average TCP
throughput at each of the 10 locations. We consider both orientation
\#0, which represents the case when both the Tx and Rx antenna arrays
are fully aligned, and the average across 16
orientations. 

Figures \ref{fig:loc_rssi} and \ref{fig:loc_xput} show that
orientation \#0 provides near best possible performance (RSSI between
80 and 100, TCP throughput between 800 and 900 Mbps) at all locations,
except one (Location \#7). In fact, the standard deviations are
negligible, indicating that the mean RSSI and throughput were
sustained across multiple runs. Location \#7 is a rather special case,
where the Tx and Rx are placed around the edges of a corner, in a
manner that there was no LOS path possible between them. Hence,
similar to the findings in~\cite{tie:pam2012}, we observe that {\em
  high-throughput 60GHz links can be established through materials
  such as walls or glass}. Although the signal does attenuate when it
passes through such materials (Figure~\ref{fig:loc_rssi}),
Figure~\ref{fig:loc_rate} shows that, in the case of optimal antenna
orientation, an NLOS link through a wall was able to sustain rates of
1540-3080 Mbps 80\% of the time and an NLOS link through glass was
able to sustain a rate of 2310 Mbps 95\% of the time.

In contrast, the performance averaged across all orientations is much
lower than for Orientation \#0; RSSI (Figure~\ref{fig:loc_rssi}) and
TCP throughput (Figure~\ref{fig:loc_xput}) never cross their halfway
mark (50 or 400 Mbps, respectively). Among setups with similar
environment but different Tx and Rx distances (Locations \#0 and \#1,
\#3 and \#4, \#8 and \#9), the setup with the smallest distance always
performs better. Further, the extremely large standard deviations
suggest very large performance variation at a given location for
different orientations. This can be attributed to some orientations
resulting in zero throughput, not even allowing a connection
establishment between the sender and receiver. For example, in the
presence of a wall or a corner between the sender and the receiver,
non-zero throughput was achieved only at 3 orientations each. Even
worse, in the case of Location \#9 (a relatively long link in a
research lab filled with ``clutter''~\cite{anderson:twc2004}, i.e.,
objects that do not directly block the LOS between the Tx and the Rx,
such as office furniture, soft partitions that do not extend to the
ceiling, and lab equipment), Orientation \#0 was able to sustain high
data rates (1925 Mbps or higher for 85\% of the time) and high
throughput, but no link was established for any of the remaining 15
orientations. Although~\cite{anderson:twc2004} found that attenuation
due to clutter decreases as we move from 2.5 GHz to 60 GHz, our
results show that clutter can have a severe impact on 60 GHz
performance, except in the case of very short distances or perfect
antenna orientation.

\noindent{\bf Remarks} Overall, these results confirm that
high-throughput 60GHz communication is feasible at various locations
typical of an indoor WLAN environment but strongly point to the
importance of the relative orientation between the Tx and Rx antenna
arrays. Although communication is possible in the case of antenna
array misalignment, either via beamsteering or through an NLOS link
via reflection, the performance can vary significantly depending on
the location and Tx-Rx distance and can be much lower than in the case
of optimal orientation.

\subsection{Impact of Tx/Rx orientation}
\label{sec:or}

\begin{figure*}[ht]
\centering 
  \subfigure[RSSI.] {
  \label{fig:or_rssi}
  \includegraphics[width=2.2in,height=1.6in]{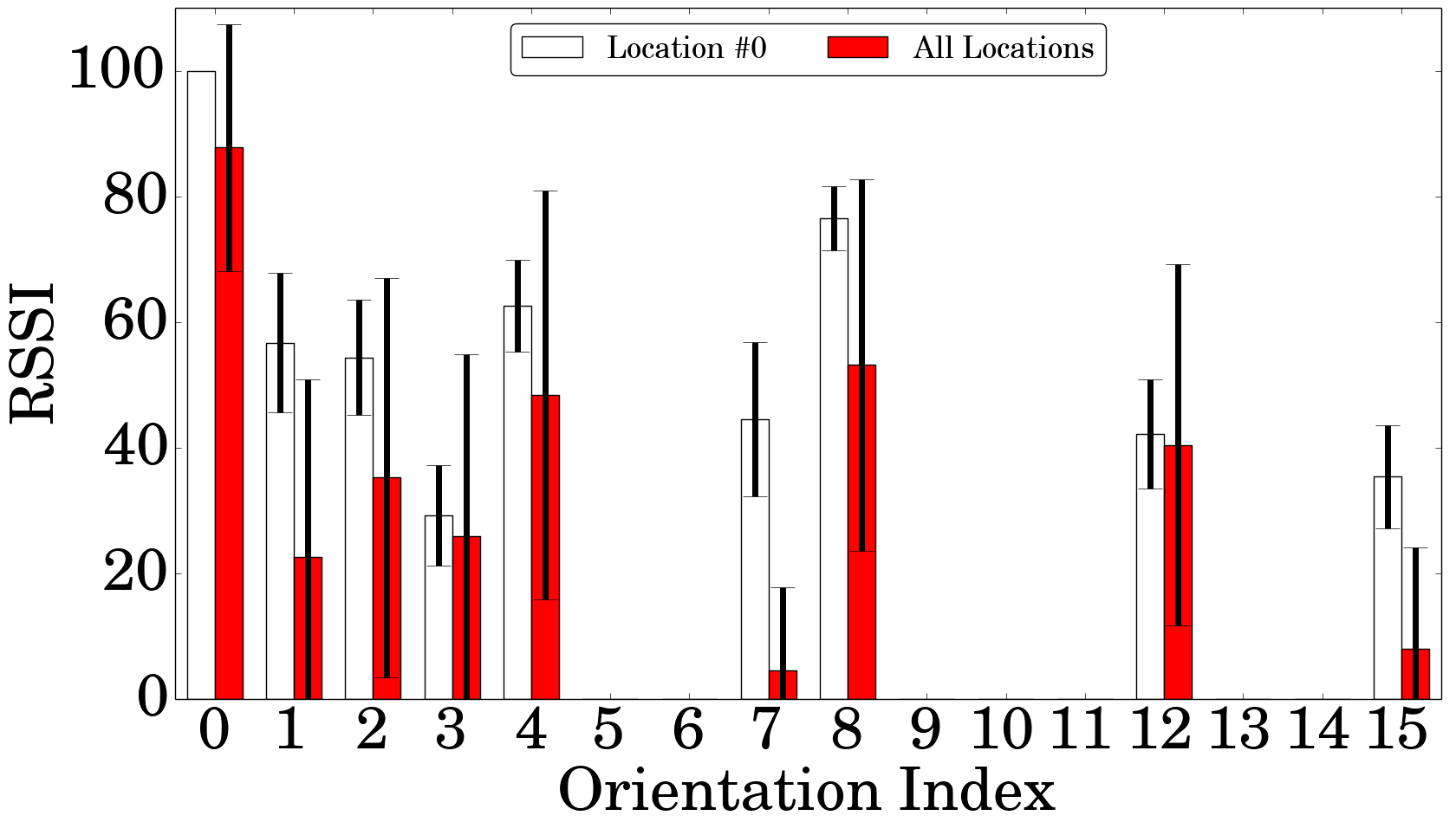} 
}
   \hspace{0.01in}
\subfigure[PHY data rate distribution.] {
  \label{fig:or_rate}
  \includegraphics[width=2.2in,height=1.6in]{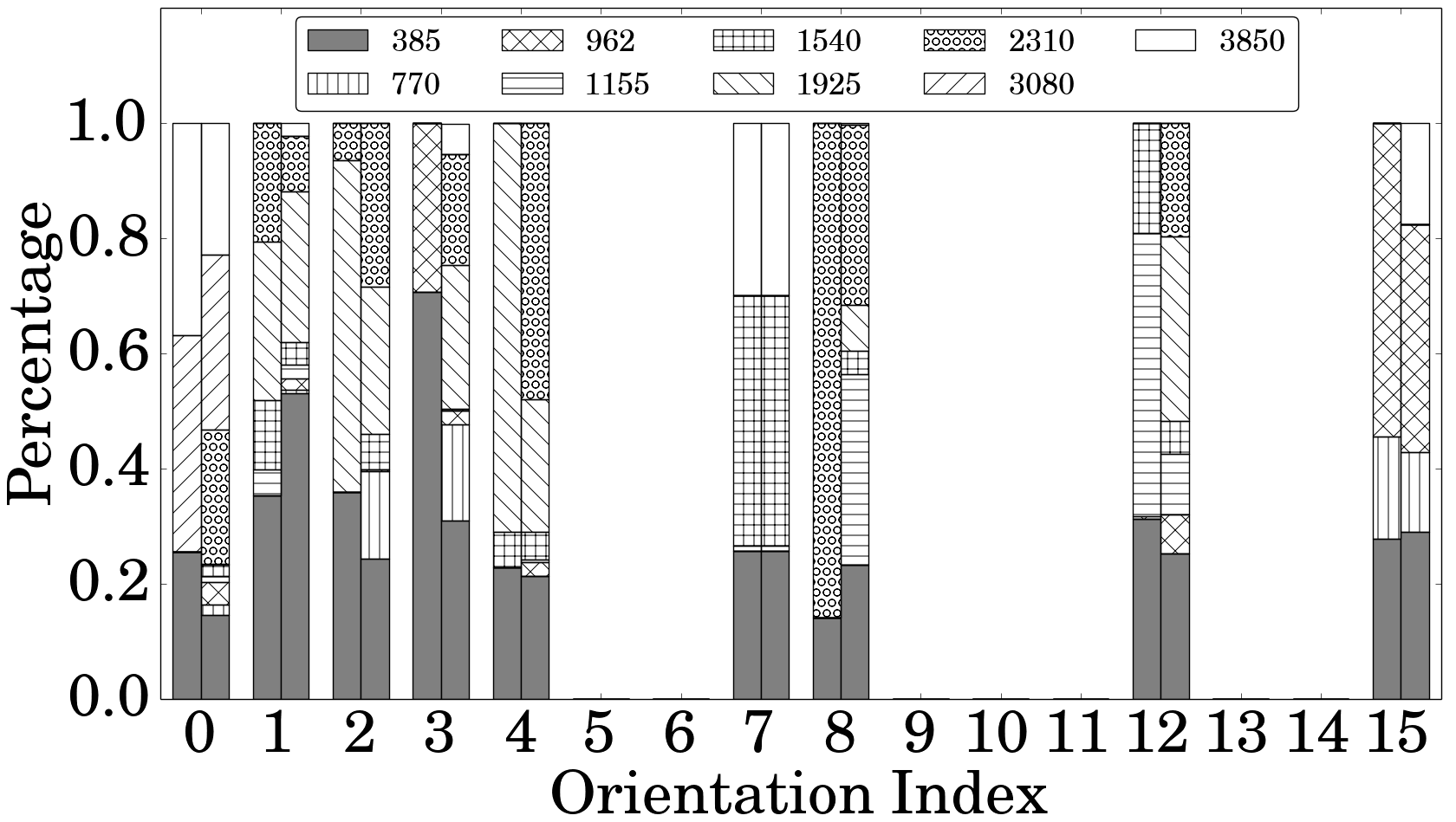}
} 
 \hspace{0.01in}
\subfigure[TCP throughput.] {
  \label{fig:or_xput}
  \includegraphics[width=2.2in,height=1.6in]{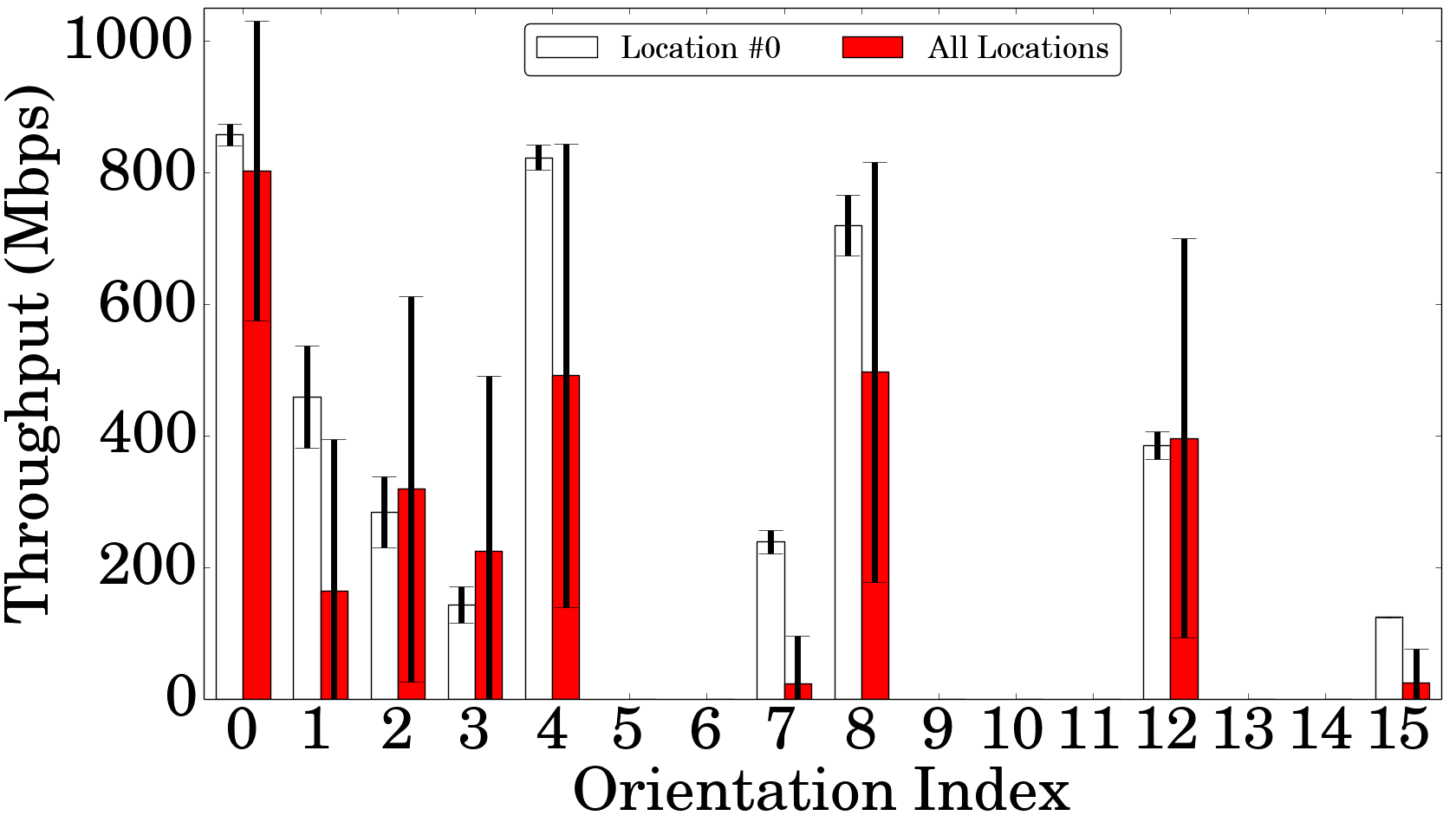} 
 }
\vspace{-0.15in}
\caption{Performance across different orientations}
\vspace{-0.15in}
\label{fig:orient}
\end{figure*}

We now take a closer look at the performance of each of the 16
orientations. Figures~\ref{fig:or_rssi},
\ref{fig:or_rate},~\ref{fig:or_xput} plot the the average RSSI, the
selected PHY rates, and the average TCP throughput at each orientation,
averaged across all locations. We also consider separately
Location\#0, which represents the best-case scenario (LOS over 8 ft).

Figure~\ref{fig:orient} shows that orientations \#0, as expected,
performs the best among all orientations. Interestingly, orientations
\#4, \#8, and \#12, i.e., cases where the Tx points directly towards
the Rx location, as in the best scenario, but the Rx is rotated by
$90^{o}$, $180^{o}$, or $270^{o}$, give very similar and significantly
higher throughput than all other orientations, indicating that the Tx
position is more critical to performance. On the other hand,
orientations \#1, \#2, and \#3 where the Rx is fixed facing the Tx
location and the Tx is rotated in $90^{o}$ intervals are characterized
by throughputs lower than 450 Mbps, and rather large standard
deviations. Even worse, for any given Tx orientation except the one
directly facing the Rx location (\#0, \#4, \#8, \#12), all Rx
orientations except the one directly facing the Tx location give
extremely low or zero performance. E.g., consider orientations \#1,
\#5, \#9 and \#13, where Tx orientation is fixed, in
Figures~\ref{fig:or_rssi},~\ref{fig:or_xput}; among them, only
orientation \#1 gives non-zero RSSI/throughput.

The heavy impact of antenna orientation may initially sound
counter-intuitive for Wilocity radios, since they are equipped with
electronically steerable antenna arrays. However, practical 802.11ad
phased antenna arrays cannot generate homogeneous beams across all
directions~\cite{rappaport:book2014}; this has been recently verified
experimentally in~\cite{sur:sigmetrics2015}.

\noindent{\bf Orientation/link asymmetry} One additional observation
from Figures~\ref{fig:or_rssi}, \ref{fig:or_rate}, \ref{fig:or_xput}
is that, at Location \#0 (Hall), orientations \#4 and \#12, which are
symmetric w.r.t the Tx position, do not give similar throughput. The
same observation can be made about orientations \#1 and \#3, which are
symmetric w.r.t the Rx position. To eliminate the impact of
environmental asymmetry (there are still walls in the Hall although
far from the Tx-Rx link, as well as furniture), we looked at the
results at Location \#2 (a Corridor with walls of the same material on
both sides). The result was similar (we omit it here due to space
limitation). Further,~\cite{sur:sigmetrics2015} showed that 60 GHz
links exhibit link asymmetry (downlink and uplink throughput are
different when the Tx and Rx use different beamwidths). These
orientation and link asymmetries make it extremely hard to predict
and/or accurately model performance in indoor environments.

\noindent{\bf Outdoor experiments} In case of Tx/Rx antenna array misalignment
(orientations other than \#0), communication can be achieved either
via beamsteering which allows for realignment of the main beam or over
an NLOS link through reflection. Since the Wilocity radios neither
allow us to control the beam steering process nor provide any
information about it, we resorted to outdoor experiments in order to
obtain a better idea about the factor that allows communication at
different orientations. Specifically, we repeated the experiments for
all 16 orientations at an outdoor open space devoid of reflective
and/or obstructive surfaces and objects. Only four orientations (0,
4,8 \& 12), where the Tx faces the Rx gave non-zero throughput.

\noindent{\bf Remarks} The results show that COTS 802.11ad devices can
establish Gigabit links even with imperfect Tx/Rx orientation. In indoor environments, the presence of multiple
reflective surfaces creates additional opportunities via NLOS links,
although the throughput of such NLOS links is typically much lower
compared to the throughput of LOS links. We conclude that Tx
orientation is more important in determining the performance and the
possibility of a connection. On the other hand, orientations where
neither of the Tx or Rx antenna point toward the other's location, are
not suitable for communication at all. Further, due to the complex
indoor environment, symmetric properties cannot be assumed.

\section{Impact of distance}
\label{sec:distance}

\begin{figure*}[ht]
\centering 
  \subfigure[RSSI.] {
  \label{fig:rssi_dist_1st}
  \includegraphics[width=2.2in,height=1.6in,]{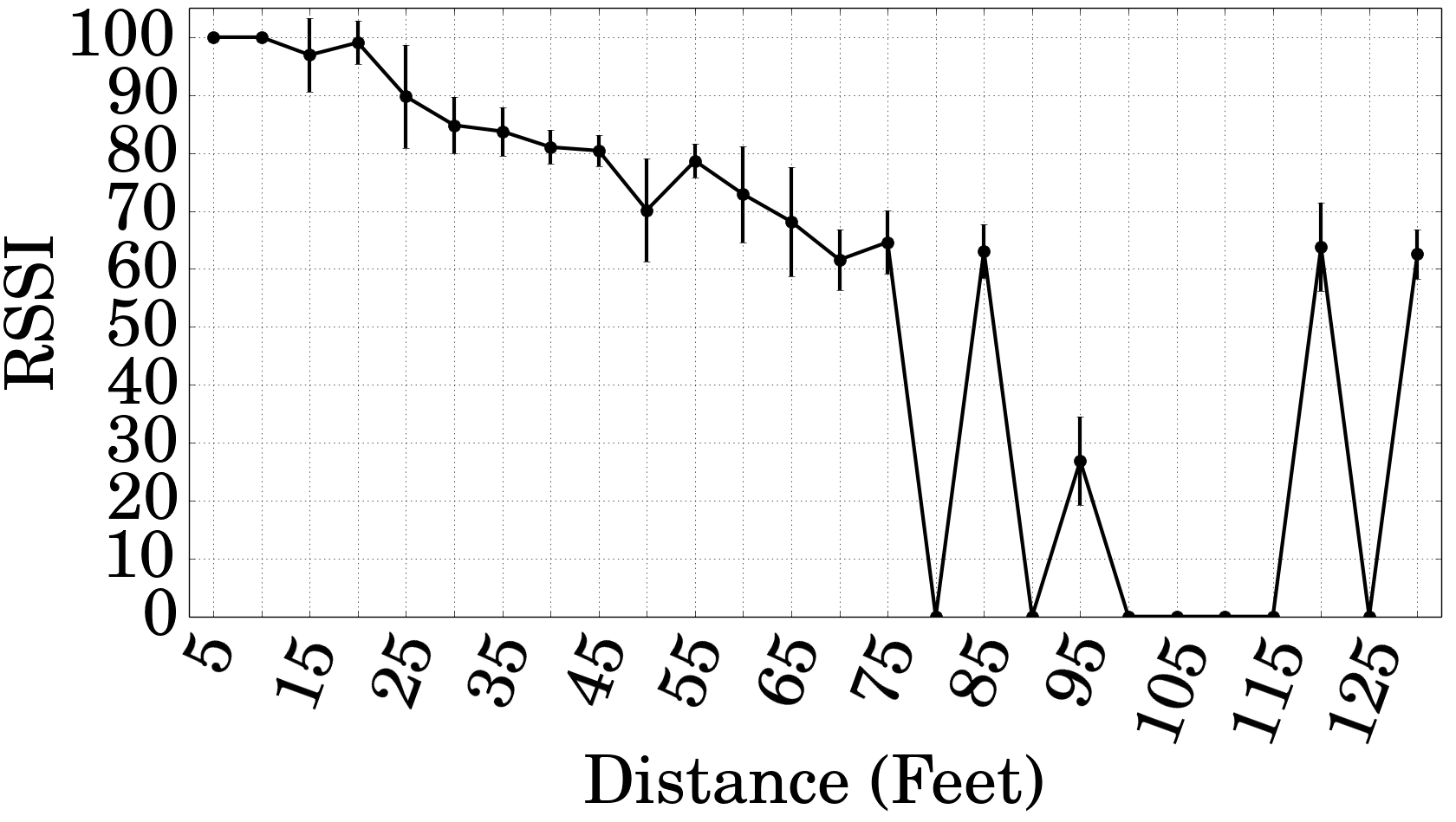} 
}
   \hspace{0.01in}
\subfigure[PHY data rate distribution.] {
  \label{fig:rate_dist_1st}
  \includegraphics[width=2.2in,height=1.6in]{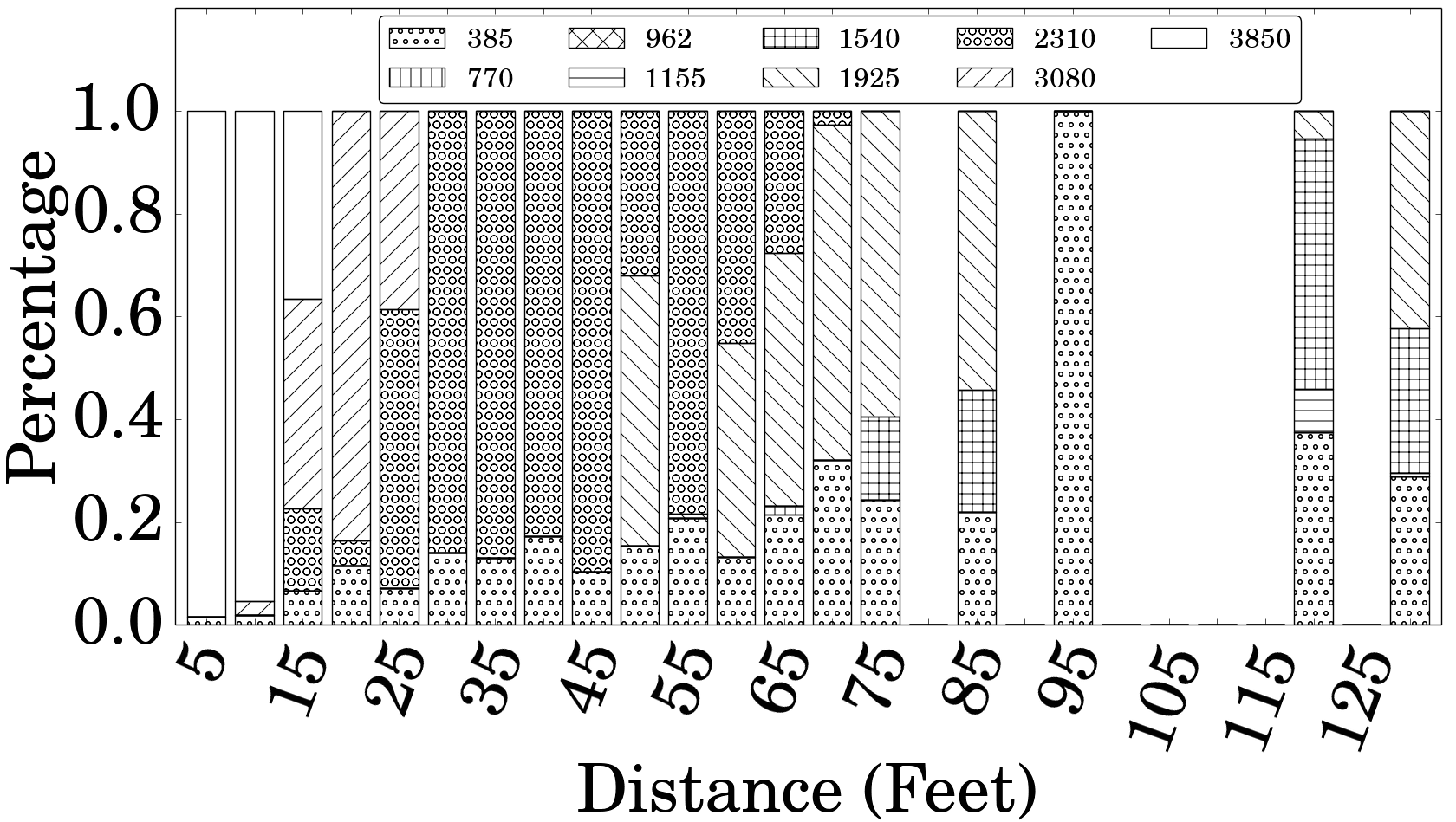}
} 
 \hspace{0.01in}
\subfigure[TCP throughput.] {
  \label{fig:xput_dist_1st}
  \includegraphics[width=2.2in,height=1.6in]{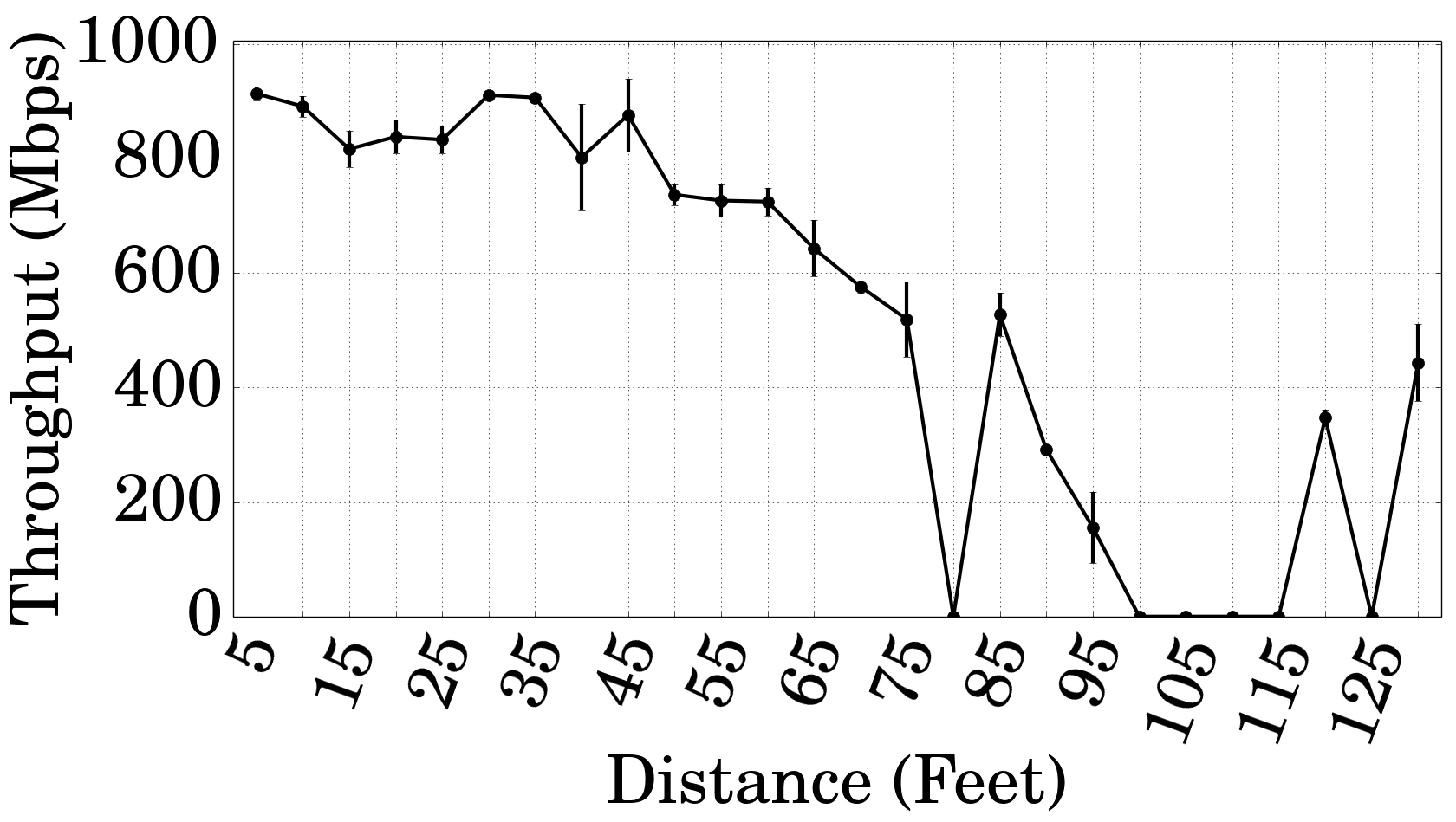} 
 }
\vspace{-0.15in}
\caption{Performance as a function of distance (Hall).}
\vspace{-0.15in}
\label{fig:dist_1}
\end{figure*}

\begin{figure*}[ht]
\centering 
  \subfigure[RSSI.] {
  \label{fig:rssi_dist_3rd}
  \includegraphics[width=2.2in,height=1.6in,]{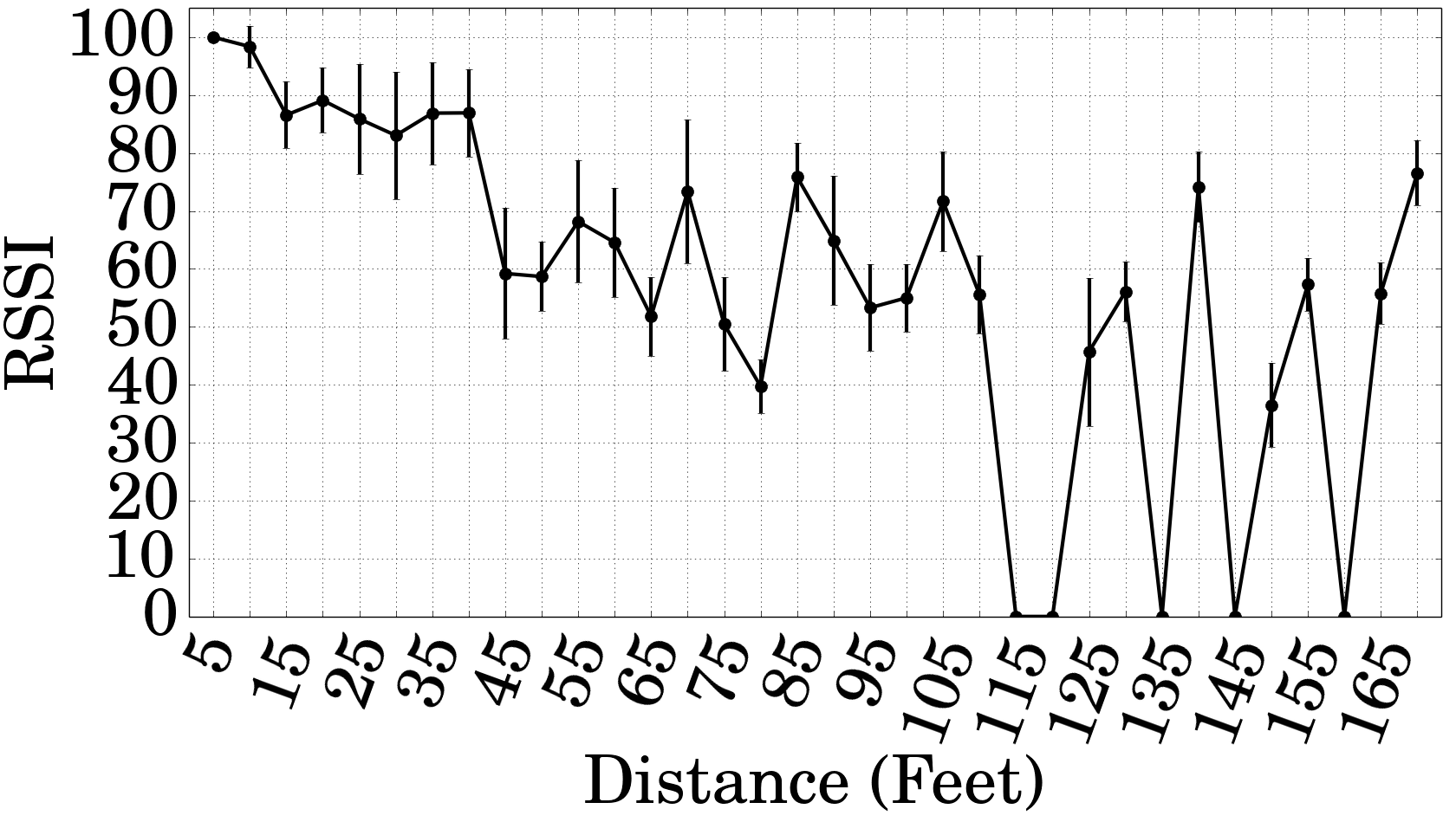} 
}
   \hspace{0.01in}
\subfigure[PHY data rate distribution.] {
  \label{fig:rate_dist_3rd}
  \includegraphics[width=2.2in,height=1.6in]{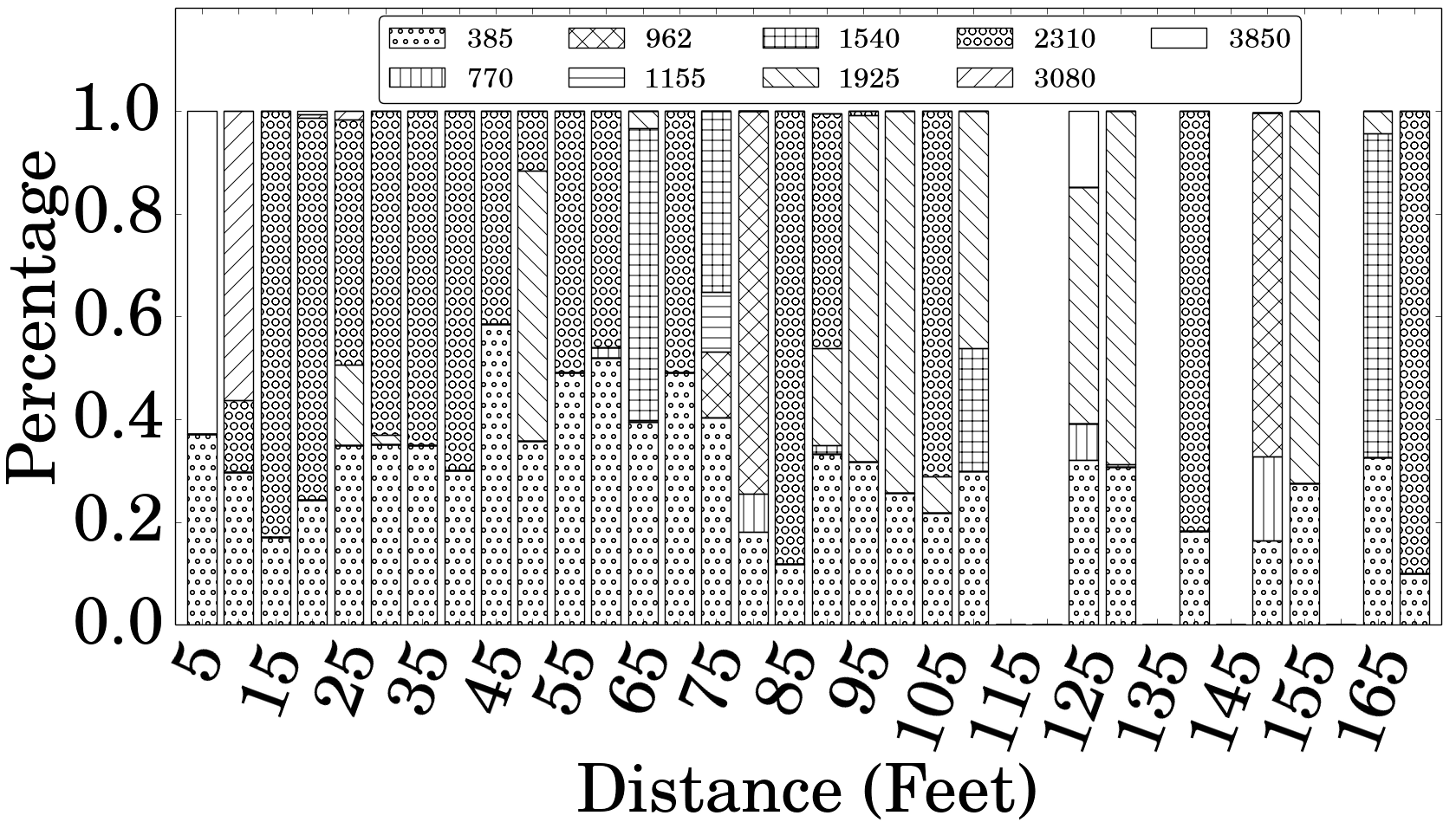}
} 
 \hspace{0.01in}
\subfigure[TCP throughput.] {
  \label{fig:xput_dist_3rd}
  \includegraphics[width=2.2in,height=1.6in]{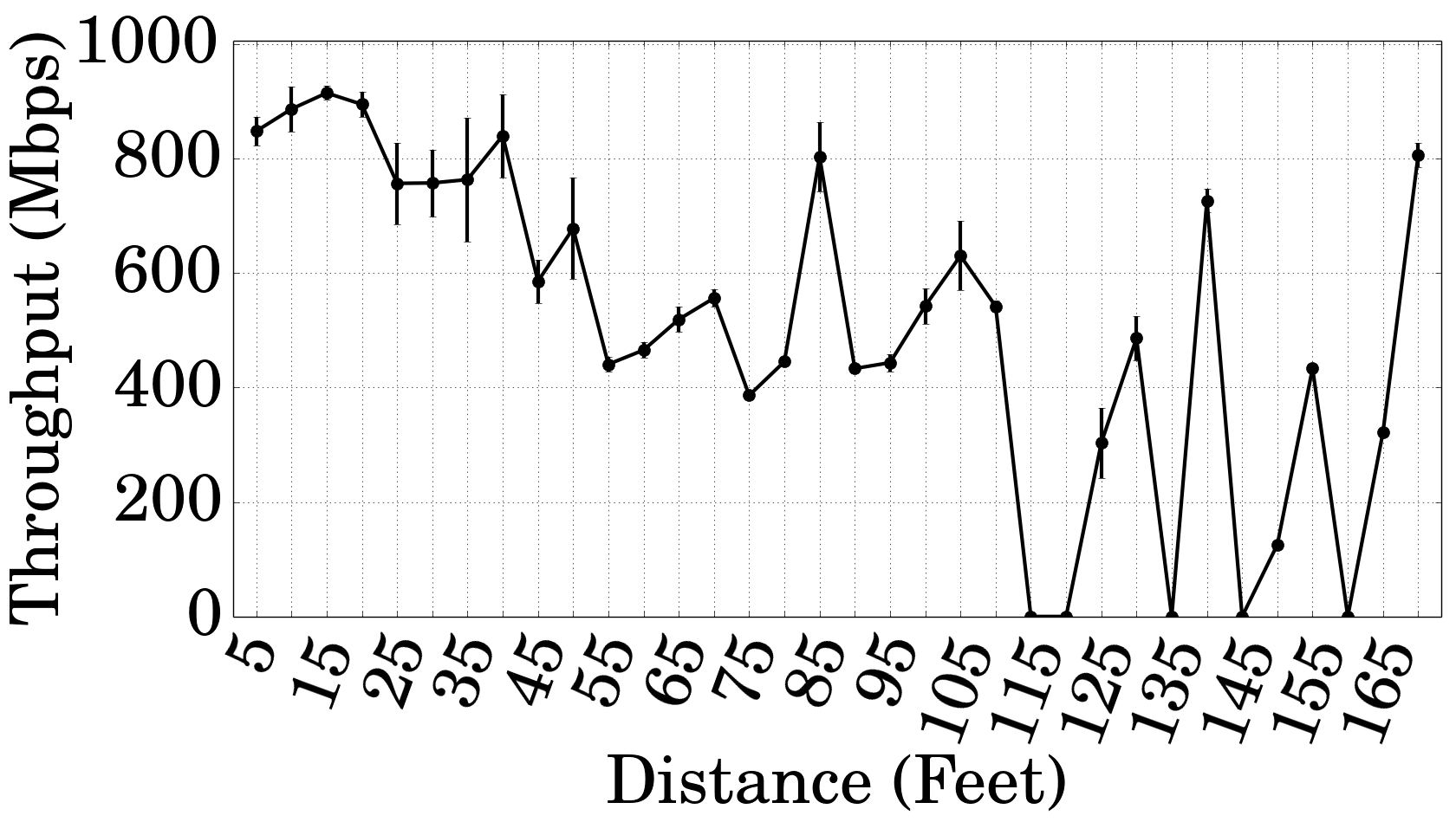} 
 }
\vspace{-0.15in}
\caption{Performance as a function of distance (Corridor).}
\vspace{-0.15in}
\label{fig:dist_3}
\end{figure*}

In the last section, we only focused on the relative orientation of Tx
and Rx antenna arrays and hence most experiments were conducted with
Tx and Rx very close to each other, to avoid the effects of severe
signal attenuation of the 60 GHz radio signal.  However, for mmWave
technology to be a viable option for building WLANs, it is necessary
to examine how channel quality indicators, RSSI and PHY data rate, and
the corresponding transport layer throughput, vary with the distance
between the transmitter and the
receiver. Figures~\ref{fig:dist_1},~\ref{fig:dist_3} plot the RSSI,
the PHY rate distribution, and the TCP throughput over distance at the
two main locations: Hall and Corridor.

\noindent{\bf Range} A common belief is that 60 GHz's range is too short even
in free space due to the very short wavelength and highly directional
antennas are required for communication at longer
distances. Figures~\ref{fig:dist_1} and~\ref{fig:dist_3} disprove this
belief, showing that long ranges can be achieved even with radios
designed for short-range indoor applications which use relatively wide
beams and have lower EIRP (23 dBm) than the maximum allowed by FCC (40
dBm)~\cite{zhu:mobicom2014}. The Corridor measurements show that RSSI
exhibits large oscillations\footnote{Due to a phenomenon known as {\em
    waveguide effect}~\cite{smulders:tap2009}.} but does not drop with
distance beyond 40 ft (Figure~\ref{fig:rssi_dist_3rd}) and a PHY data
rate of 2310 Mbps can be supported at a distance of 170 ft
(Figure~\ref{fig:rate_dist_3rd}). The Hall measurements show a
different picture, closer to what one would expect, with RSSI
gradually dropping with distance up to 75 ft
(Figure~\ref{fig:rssi_dist_1st}) but even in this case, the link was
able to support a rate of 1540 Mbps or 1925 Mbps roughly 70\% of the
time at a distance of 130 ft (Figure~\ref{fig:rate_dist_1st}). These
ranges are much longer than the values reported recently with the same
hardware (770 Mbps at 72 ft in a datacenter~\cite{zhu2:mobicom2014},
385 Mbps at 72 ft and 2310 Mbps at only 33 ft in an outdoor
environment~\cite{zhu:mobicom2014}).

\noindent{\bf RSSI vs. distance} Recent experimental
work~\cite{halperin:sigcomm2011,zhou:sigcomm2012,zhu:mobicom2014,zhu2:mobicom2014}
observed that the attenuation of 60 GHz signals with distance follows
closely the Friis model in LOS scenarios, both in stable datacenter
and outdoor picocell environments. The validity of the free-space
propagation model has led to the use of simple RSS-based rate
adaptation algorithms in simulators
~\cite{halperin:sigcomm2011,zhou:sigcomm2012,tie:pam2012,zhu2:mobicom2014}
and the use of RSS as a direct indicator of the PHY data
rate~\cite{zhu:mobicom2014,sur:sigmetrics2015}. We discuss the
validity of the Friis model in indoor WLAN environments in this
section and the relationship between PHY data rate and RSSI in
Section~\ref{sec:predict}.

In Figure~\ref{fig:dist_1} (Hall), we observe that the distance axis
can be divided in 3 distinct regions. For distances up to 20 ft, RSSI
remains close/equal to 100 and the link can sustain the highest PHY
data rate almost 100\% of the time (Figure~\ref{fig:dist_1}). The next
region is between 25 ft and 75 ft where RSSI decreases with
distance. Lastly, distances between 80 ft and 130 ft are characterized
by extremely large RSSI oscillations; RSSI drops to zero at several
distances and then rises again, often to high levels. Although we
cannot confirm it, we believe these link outages are the result of
multipath. We also hypothesize that such ``dead zones'' might have led
researchers previously~\cite{zhu:mobicom2014,zhu2:mobicom2014} to
conclude a much shorter range for the Wilocity radios. It is possible
that narrower beams can eliminate dead zones at the cost of higher
vulnerability to blockage and
mobility\cite{sur:sigmetrics2015}. Investigating this tradeoff is part
of our future work.

In Figure~\ref{fig:dist_3} (Hall), we observe a different behavior,
characterized by 4 distinct zones. RSSI shows a decreasing trend with
distance only for very short distances (5-15 ft), remains almost
stable for distances 20-40 ft, exhibits very large variations and
non-monotonic behavior but non-zero values for distances longer 40-110
ft, and finally exhibits ``dead zones'' at distances longer than 110
ft.

Overall, {\em the large variability of RSSI with distance indicates
  the presence of strong multipath in typical WLAN environments.}

\noindent{\bf PHY data rate vs. distance} Figure~\ref{fig:rate_dist_1st} and
\ref{fig:rate_dist_3rd} show that for most distances there are 2 or 3
dominant data rates, and the lowest rate of 385 Mbps is used at least
10\% and up to 60\% of the time, even in the case of very short
distances/high RSSI (with the exception of very short distances in the
Hall). {\em This observation suggests highly time-varying channels
  and/or inability of the rate adaptation algorithm to converge to a
  single rate} (note that the experiments were performed in a
stable environment with no human presence).

In the Hall experiments (Figure~\ref{fig:rate_dist_1st}), we still
observe a monotonic decrease with distance and RSSI; lower data rates
dominate at longer distances/lower RSSI values. In contrast, the is no
such monotonicity in the corridor.

\begin{figure}[ht]
\centering 
  \subfigure[Hall.] {
  \label{fig:domrate_vs_dist_1st}
  \includegraphics[width=1.5in,height=1.1in,]{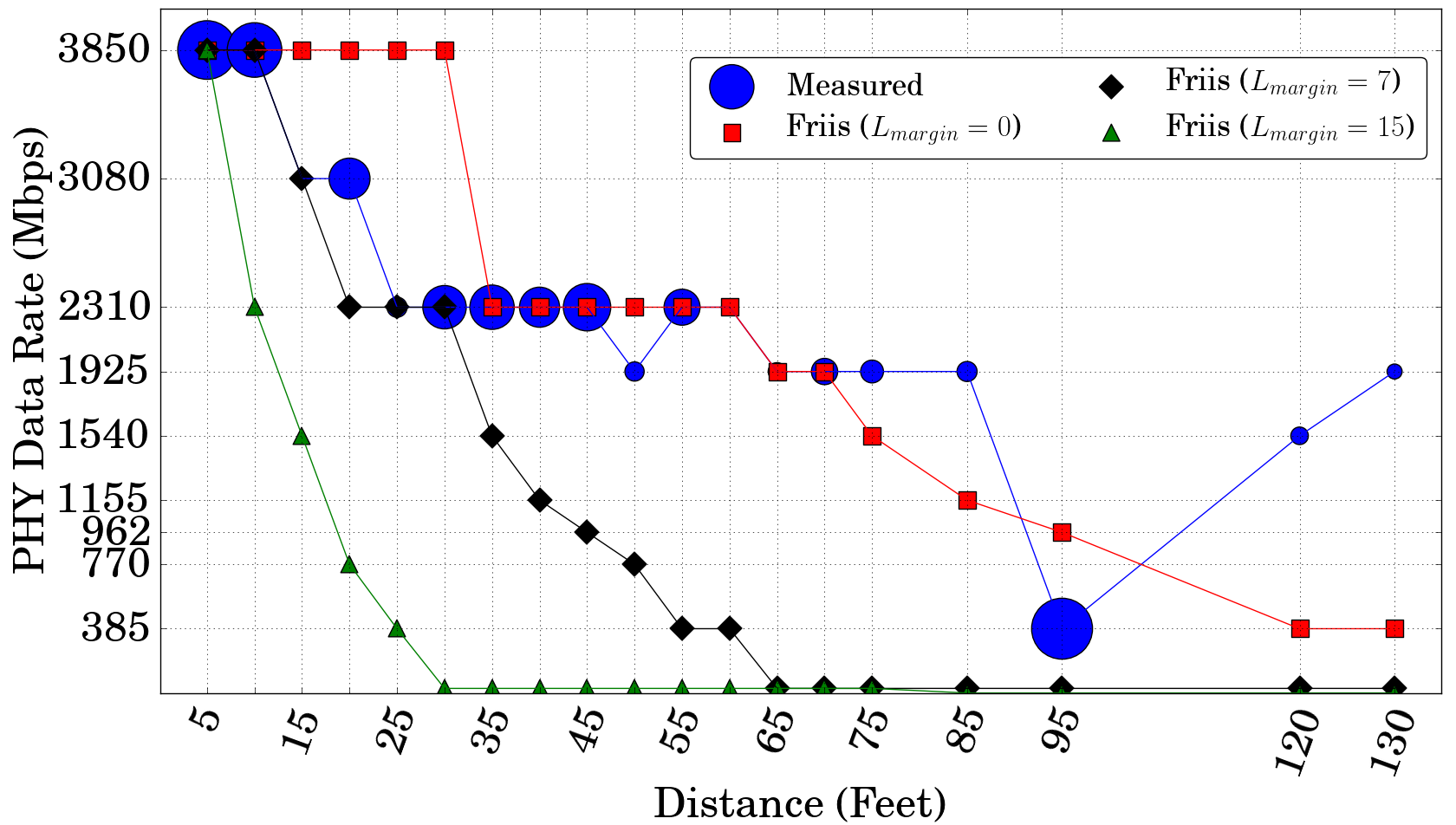} 
}
   \hspace{0.01in}
\subfigure[Corridor.] {
  \label{fig:domrate_vs_dist_3rd}
  \includegraphics[width=1.5in,height=1.1in]{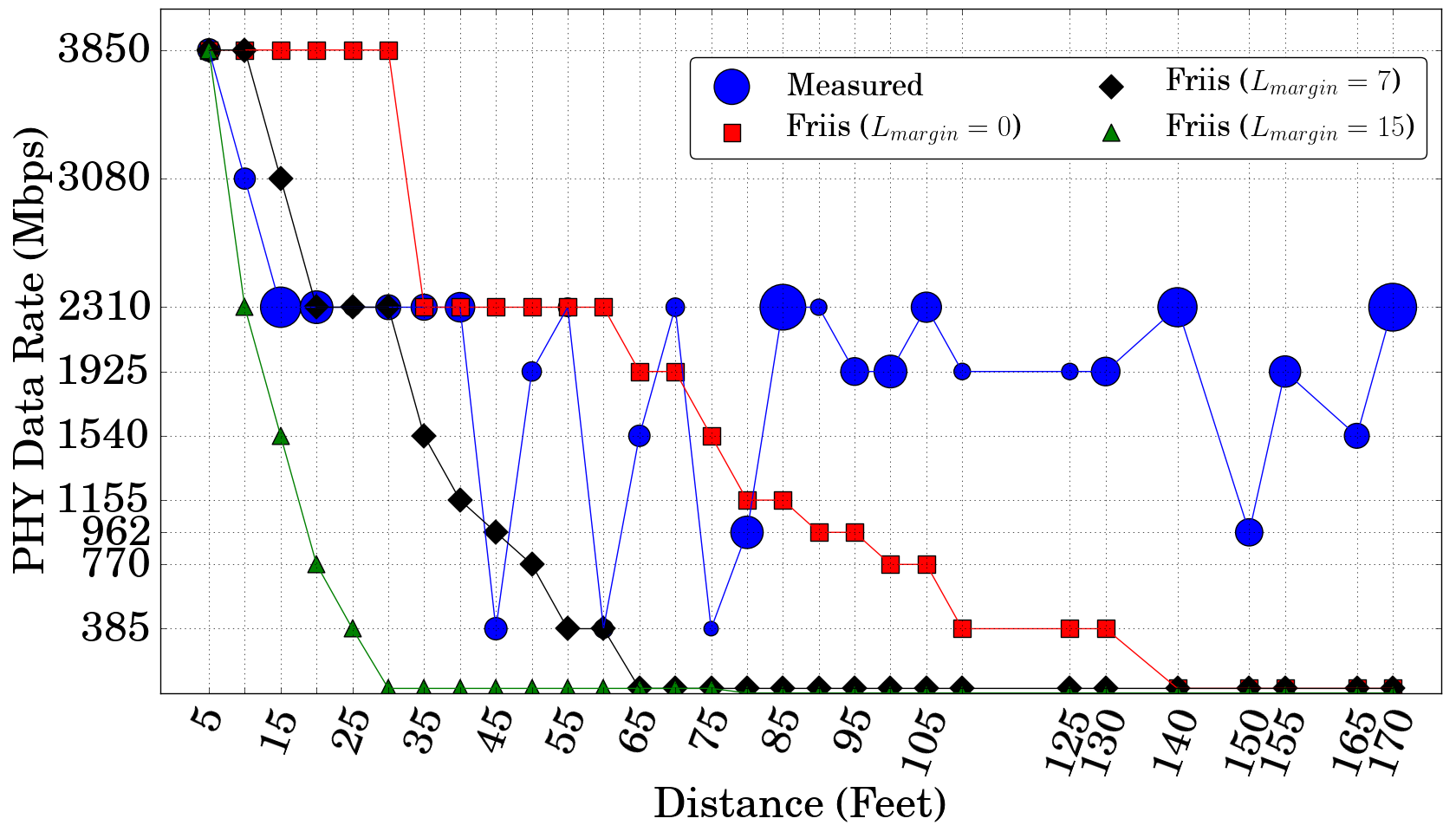}
} 
\vspace{-0.15in}
\caption{Comparison of measured dominant rate vs. the theoretically
  calculated rate from the log distance path loss model and the
  802.11ad rate-sensitivity table. In the case of the dominant rate
  curves, a larger circle indicates larger dominance.}
\vspace{-0.15in}
\label{fig:domrate}
\end{figure}

As mentioned previously, the assumption of Friis propagation model (or
more generally a log distance path loss model) has led to the use of
simple RSS-based rate adaptation algorithms in simulators
~\cite{halperin:sigcomm2011,zhou:sigcomm2012,tie:pam2012,zhu2:mobicom2014}
and the use of RSS as a direct indicator of the PHY data
rate~\cite{zhu:mobicom2014,sur:sigmetrics2015}. Our results in Figures
\ref{fig:rssi_dist_1st} and \ref{fig:rssi_dist_3rd} clearly showed
that in indoor WLANs RSSI does not decrease monotonically with
distance. Since we cannot directly compare the measured propagation
characteristics with those of the Friis model due to the fact that our
cards do not report the received signal strength (RSS), we now attempt
an indirect comparison via the supported PHY data rates. Specifically, for each
distance we calculate a theoretical RSS value $P_{RX}$ based on the commonly
used log-distance path loss model adjusted to account for shadowing
fading~\cite{smulders:tap2009} and potential losses due to reflections
in case of NLOS links~\cite{zhu:mobicom2014}, as follows: 

\begin{equation}
\begin{split}
P_{RX}(dBm) = EIRP(dBm) + G_{RX}(dBi) - \\ 
- L_{ploss}(R) + X_{\Omega} - L_{margin} \label{eqn:1} 
\end{split}
\end{equation}
\begin{equation}
L_{ploss}(R) = 10log_{10}\frac{16\pi^2R^2}{\lambda^n} \label{eqn:2}
\end{equation}

\noindent where $EIRP$ is equal to $23 dBm$ for Wilocity
Radios~\cite{zhu:mobicom2014}, $G_{RX}(dBi) = 10log_{10}N_{RX}$ is the
receiver antenna gain as a function of the antenna elements $N_{RX} =
16$~\cite{zhu:mobicom2014}, $n$ is the path loss exponent (we use
different values for Corridor and Hall based
on~\cite{smulders:tap2009}) and $X_{\Omega}$ represents a shadowing
component (zero mean Gaussian random variable with standard deviation
values also obtained from~\cite{smulders:tap2009} for different
environments). In~\cite{zhu:mobicom2014}, $L_{margin}$ is taken equal
to 15 dB although most materials typically lead to 6-7 dB loss. In our
case, we consider three different values -- 0, 7, and 15 dB. We then
use the rate-sensitivity table for 802.11ad (Table 2
in~\cite{sur:sigmetrics2015}) to convert $P_{RX}$ to a PHY data rate.

Figures \ref{fig:domrate_vs_dist_1st} and
\ref{fig:domrate_vs_dist_3rd} compare the measured dominant rates
against the theoretically computed rates from (\ref{eqn:1}). We
observe that the conservative models which account for reflection
losses significantly underestimate the data rate; if we assume a 15/7
dB loss, only the control data rate (27.5) Mbps can be supported for
distances longer than 30/65 ft. On the other hand, assuming zero loss
due to reflections results in overestimation of the data rate for
short distances (up to 30 ft) and underestimation for long distances
in both environments, potentially due to a combination of multipath
and waveguide effects. Overall, we observe that {\em PHY data rate
  cannot be predicted from simple propagation models in indoor WLAN
  settings}. 

\noindent{\bf Throughput vs. distance} In Figures \ref{fig:xput_dist_1st} and
\ref{fig:xput_dist_3rd} we observe again distinct regions on the
distance axis although these regions do not always overlap with the
RSSI regions. In the Hall experiments (Figure
\ref{fig:xput_dist_1st}), it sustains high values (above 800 Mbps) for
distances up to 45 ft although RSSI starts dropping at 25 ft. It then
exhibits a gradual drop up to a distance of 75 ft (boundary of the
second RSSI region) and ``dead zones'' for longer distances. In the
Corridor experiments (Figure \ref{fig:xput_dist_3rd}), we observe two
small regions of constant values (around 900 Mbps for at 5-20 ft,
around 800 Mbps at 25-40 ft), very large variations with distance (up
to 400 Mbps within 5 ft) for distances up to 110 ft, and ``dead
zones'' for longer distances. Overall, we observe {\em a weak
  correlation of throughput with distance in certain environments and
  no correlation in other environments}.

\noindent{\bf Remarks} The results in this section show that RSSI does
not drop monotonically with distance. This observation suggests that,
in contrast to observations made by previous works in indoor
datacenter or outdoor picocell environments, propagation in typical
indoor WLAN environments does not follow simple propagation models
from the literature due to the presence of strong multipath, and calls
for new propagation models in 802.11ad simulators. The rate adaptation
logic cannot converge to a single rate most of the time even in the
case of high RSSI indicating a weak (if any) correlation between the
two metrics. PHY data rate and throughput show a weak correlation with
distance in certain environments (Hall) and no correlation in other
environments (Corridor).

\section{Relationship among RSSI, PHY data rate, and throughput}
\label{sec:predict}

\begin{figure*}[ht]
\centering 
  \subfigure[PHY data rate vs. RSSI.] {
  \label{fig:rate_vs_rssi_1}
  \includegraphics[width=2.2in,height=1.6in,]{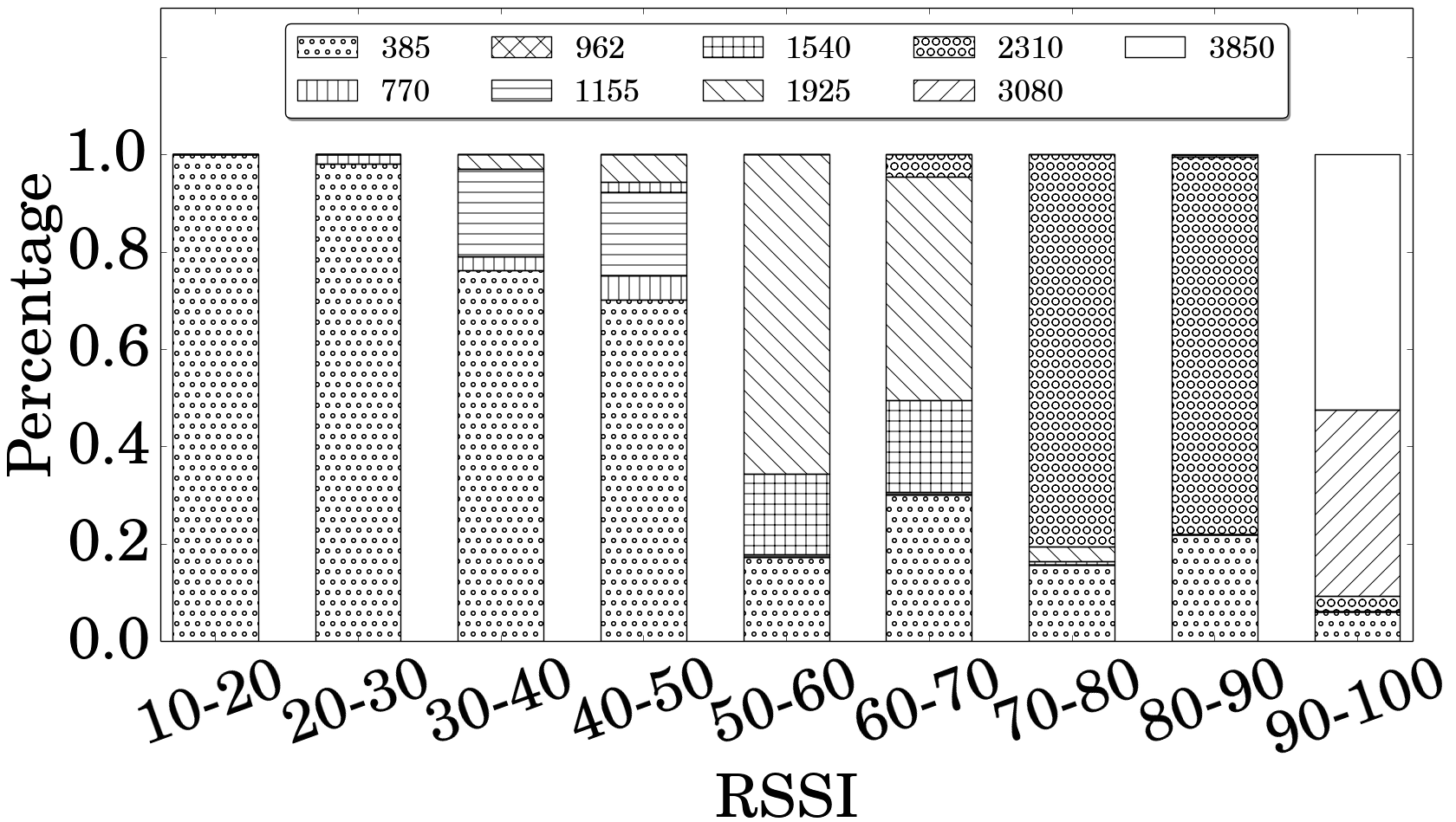} 
}
   \hspace{0.01in}
\subfigure[TCP throughput vs. RSSI.] {
  \label{fig:tput_vs_rssi_1}
  \includegraphics[width=2.2in,height=1.6in]{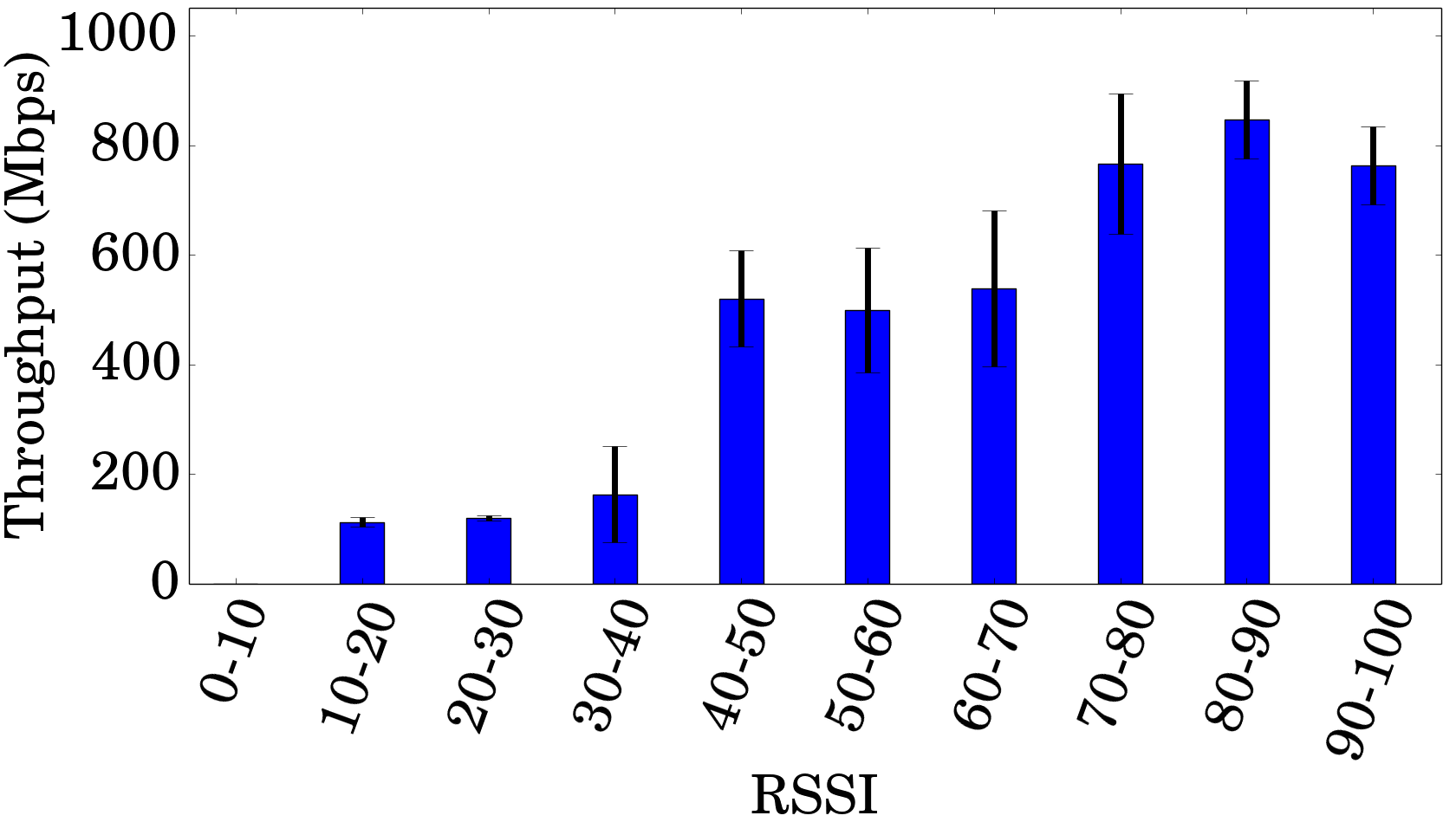}
} 
 \hspace{0.01in}
\subfigure[TCP throughput vs. PHY data rate.] {
  \label{fig:tput_vs_rate_1}
  \includegraphics[width=2.2in,height=1.6in]{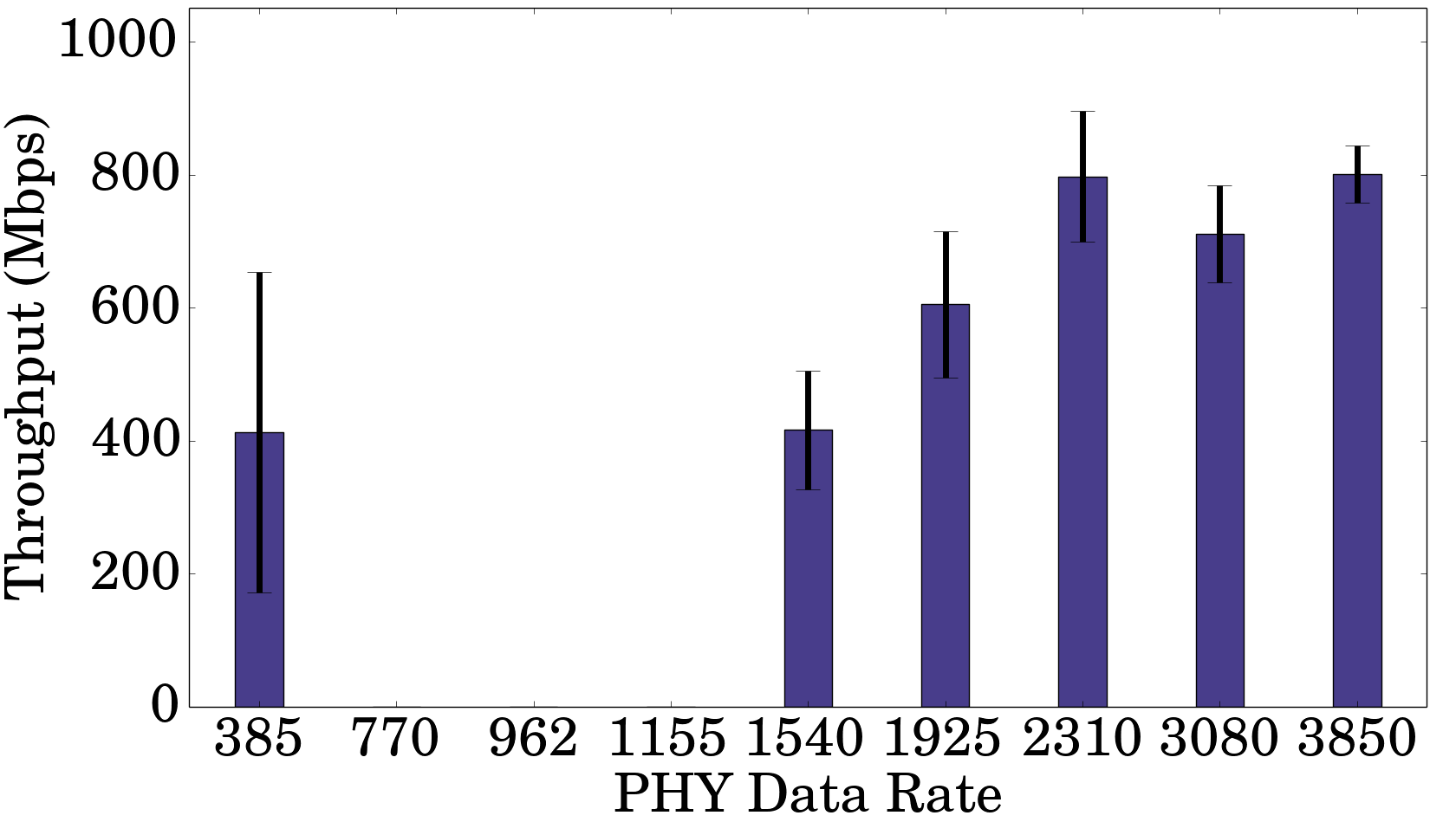} 
 }
\vspace{-0.15in}
\caption{Relationship among RSSI, PHY data rate, and throughput (Hall).}
\vspace{-0.15in}
\label{fig:predict_dist_1}
\end{figure*}

\begin{figure*}[ht]
\centering 
  \subfigure[PHY data rate vs. RSSI.] {
  \label{fig:rate_vs_rssi_3}
  \includegraphics[width=2.2in,height=1.6in,]{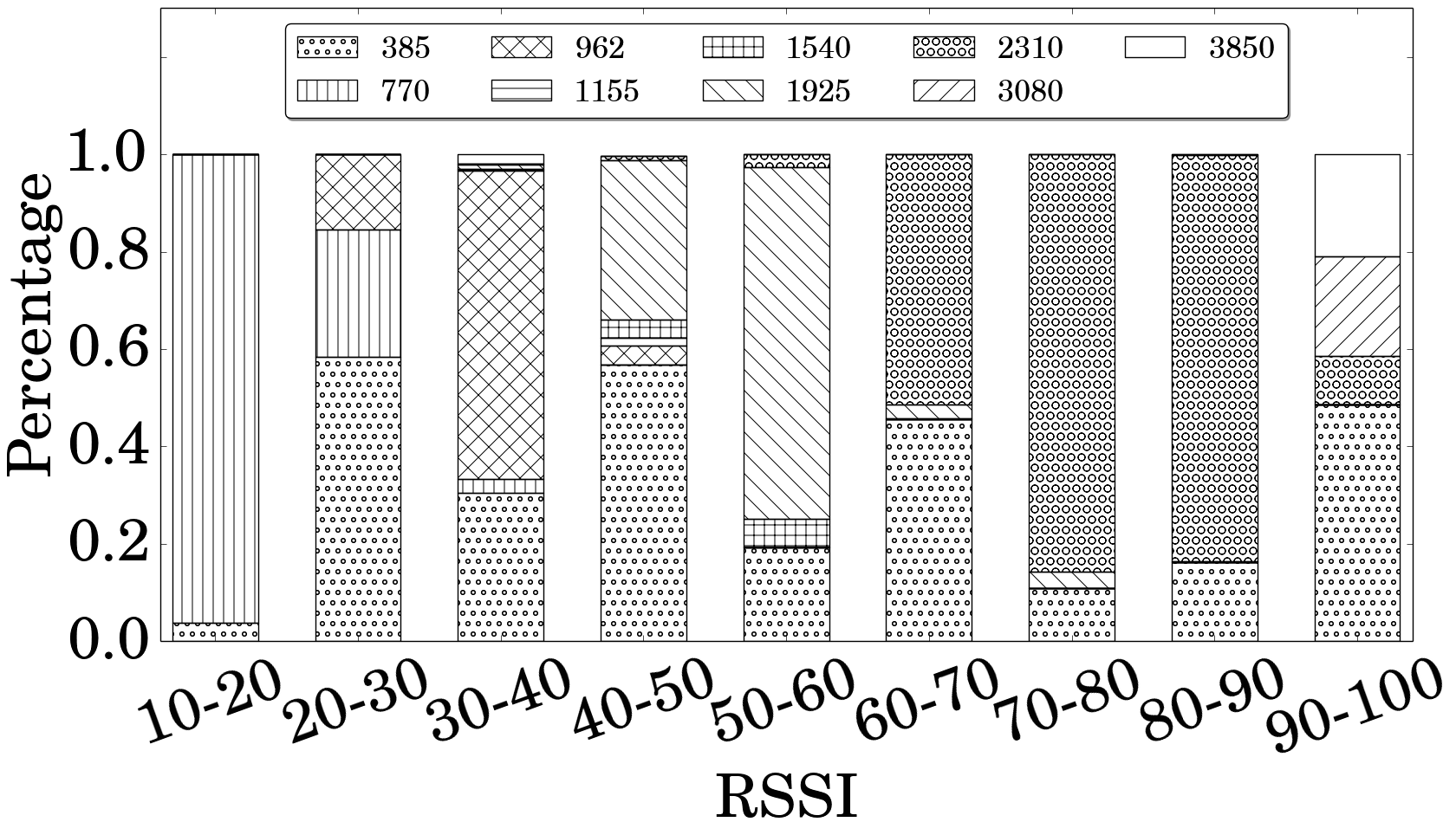} 
}
   \hspace{0.01in}
\subfigure[TCP throughput vs. RSSI.] {
  \label{fig:tput_vs_rssi_3}
  \includegraphics[width=2.2in,height=1.6in]{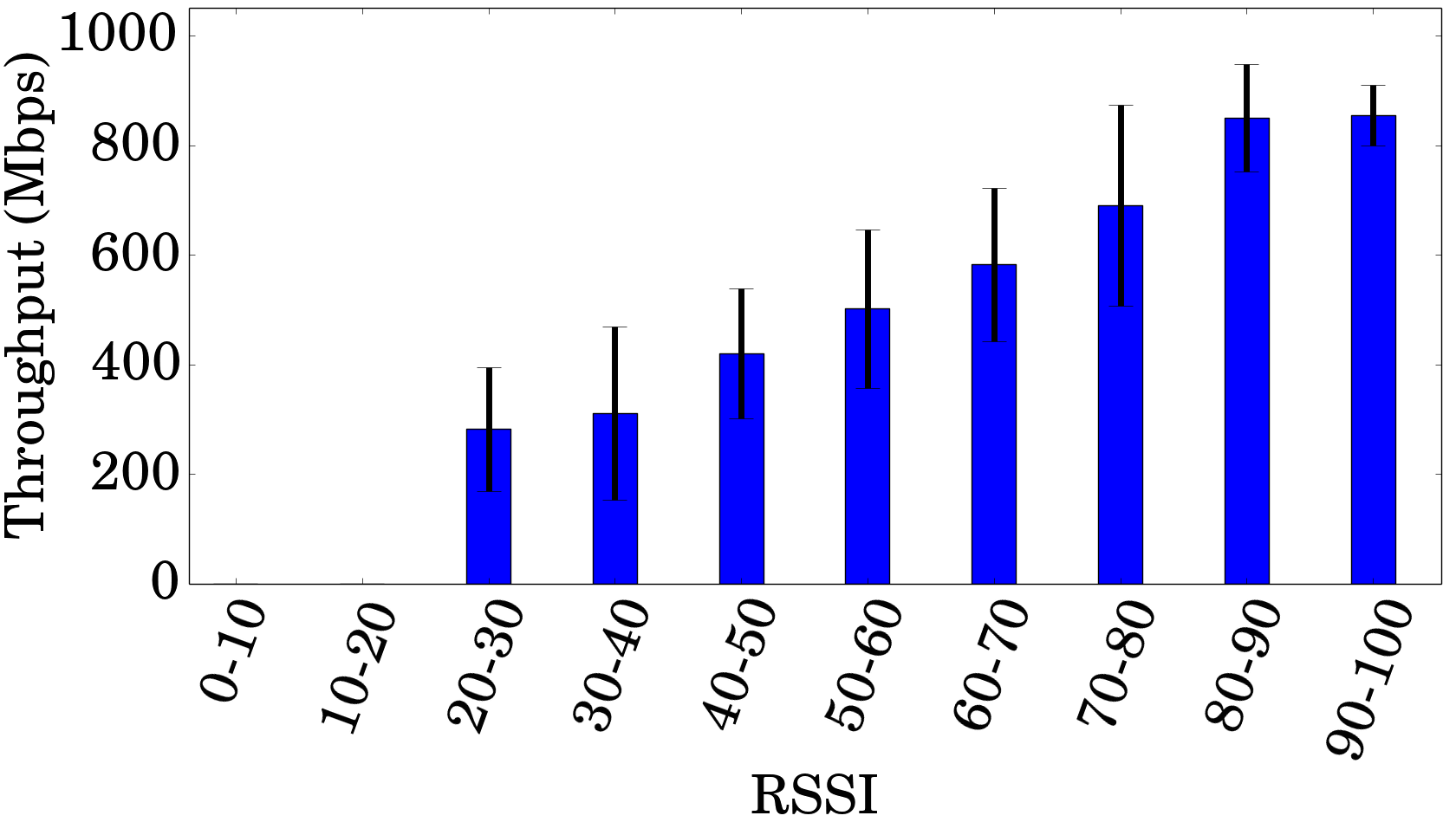}
} 
 \hspace{0.01in}
\subfigure[TCP throughput vs. PHY data rate.] {
  \label{fig:tput_vs_rate_3}
  \includegraphics[width=2.2in,height=1.6in]{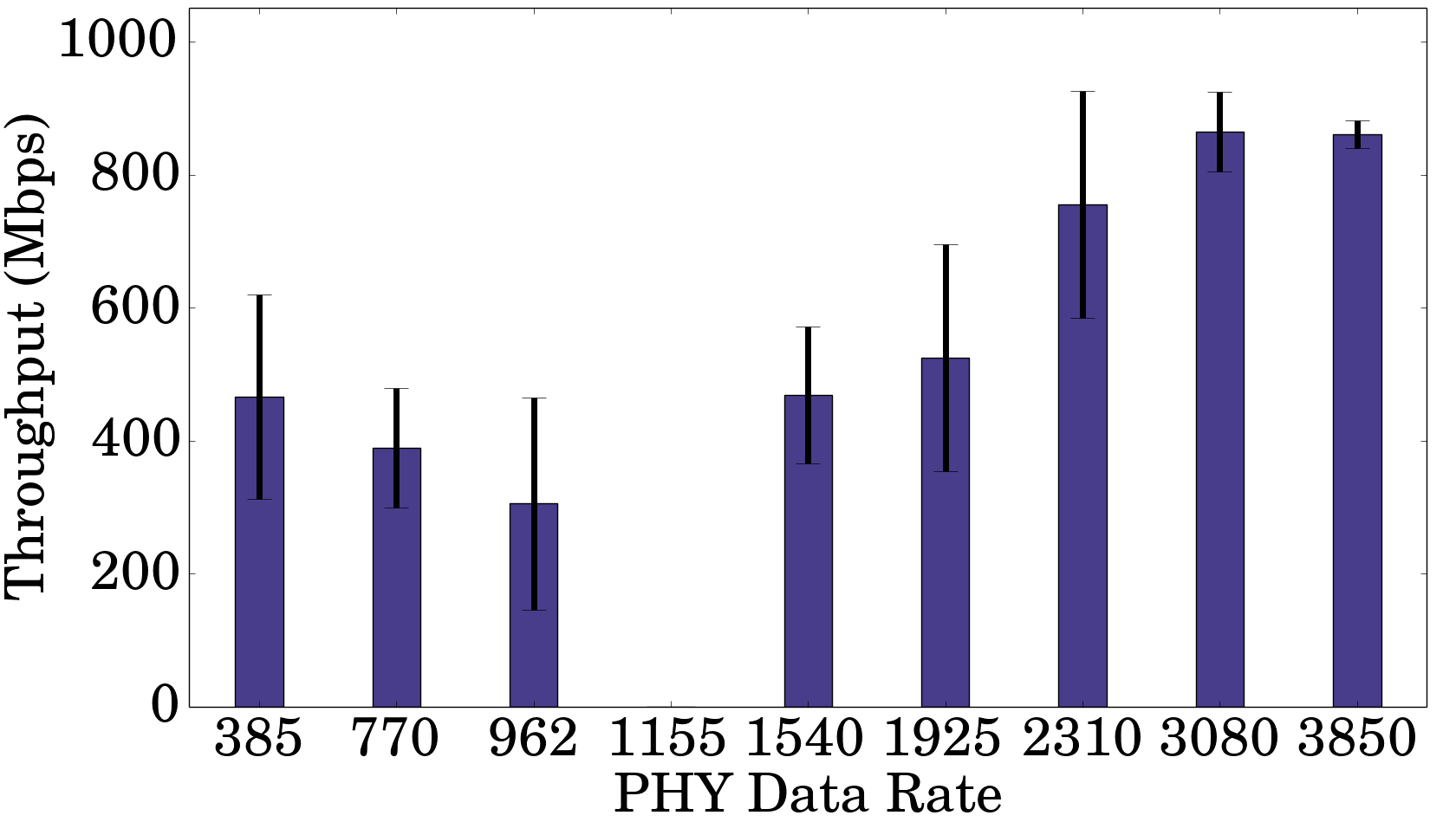} 
 }
\vspace{-0.15in}
\caption{Relationship among RSSI, PHY data rate, and throughput (Corridor).}
\vspace{-0.15in}
\label{fig:predict_dist_3}
\end{figure*}

In the previous section, we saw that distance cannot serve as a strong
indicator of link quality (RSSI, PHY data rate) and higher layer
throughput. In this section, we take a closer look among the three
metrics -- RSSI, PHY data rate, and TCP throughput -- and investigate
whether one of them can be used as a strong indicator of the other. In
particular, we examine whether (i) RSSI can be used to predict PHY
data rate and/or TCP throughput and (ii) PHY data rate can be used to
predict TCP throughput.

\noindent{\bf PHY data rate, throughput vs. RSSI} Since RSSI varies during a 10
sec iperf 3 session, we had to consider a finer time granularity to
investigate the relationship between RSSI and PHY data rate/TCP
throughput. We divided each session in 100 ms intervals and selected
only those intervals where a particular RSSI value was observed at
least 90\% of the time. We then grouped the dominant RSSI values
observed in the selected intervals in bins of 100 units. Figures
\ref{fig:rate_vs_rssi_1} and \ref{fig:tput_vs_rssi_1} plot the PHY
data rate distribution and the average TCP throughput over RSSI in the
Hall. Figures \ref{fig:rate_vs_rssi_3} and \ref{fig:tput_vs_rssi_3}
repeat the same graphs in the Corridor.

We fist focus on PHY data rate. Figures \ref{fig:rate_vs_rssi_1} and
\ref{fig:rate_vs_rssi_3} show that RSSI can serve as a (weak)
indicator of PHY data rate at a given location; for most RSSI values,
there is a dominant data rate appearing more than 60\% of the
time. The situation is better in the Hall where we also observe a
monotonic relationship between the two metrics -- higher dominant data
rates for higher RSSI values. However, the picture changes when we
compare the two locations. For the same RSSI bin, the observed data
rates can be very different at the two locations. For example, for low
RSSI (20-30), the data rate remains constant at 385 Mbps in the Hall
but can take the values of 770 Mbps or 962 Mbps 40\% of the time in
the Corridor. As another example, for very strong RSSI (90-100), the
data rate in the Corridor takes its lowest value (385 Mbps) 50\% of
the time.

We now look at the throughput. Figure~\ref{fig:tput_vs_rssi_1} shows
that RSSI can serve as a reliable although coarse-grained indicator of
throughput in the Hall. We clearly distinguish 3 regions -- high
throughput region (600-900 Mbps for RSSI higher than 70), medium
throughput region (400-700 Mbps for RSSI between 40 and 70), and low
throughput region (0-300 Mbps for RSSI lower than 40), with only one
``grey zone'' (RSSI 60-70 and 70-80). The picture is very different in
the Corridor. Instead of distinct regions, here we observe a monotonic
increase of the average throughput with RSSI. However, the standard
deviations are very large (100-200 Mbps) except in the case of very
high RSSI values. We also observe that, for the same RSSI, throughput,
similar to the data rate, can be very different in the two locations,
making prediction different across locations.

\noindent{\bf Throughput vs. PHY data rate} Similar to RSSI, the PHY data rate
also varies during a 10 sec iperf 3 session. Hence, we used a similar
methodology to investigate the relationship between PHY data rate and
TCP throughput. We selected only the 100 ms intervals where a
particular data rate was reported at least 90\% of the time. Figures
\ref{fig:tput_vs_rate_1} and \ref{fig:tput_vs_rate_3} plot the average
TCP throughput over the PHY data rate in the Hall and Corridor,
respectively.

A first observation from these figures is that some data rates are
never selected consistently over a 100 ms period. The two highest data
rates result in high throughput values and small standard deviations
in both locations. However, for the remaining data rates, throughput
varies significantly with standard deviations often higher than 200
Mbps. In the Hall, higher data rates typically result in higher
throughput ranges. On the other hand, in the Corridor, several data
rates have overlapping throughput ranges. Overall, the PHY data rate
cannot serve always as a good indicator of throughput.

\noindent{\bf Remarks} The results in this section show that RSSI can only
serve as a weak indicator of PHY data rate and TCP throughput and only
at certain locations, but not across locations. This observation has
two immediate implications. First, translating signal strength to PHY
data rate, a common practice in recent measurement
studies~\cite{zhu:mobicom2014,sur:sigmetrics2015}, can yield
inaccurate results in typical indoor WLAN environments; the same
observation is true for legacy 802.11
(e.g.,~\cite{wong:mobicom2006,pefkianakis:mobicom2010}). Second,
simple RSS-based rate adaptation algorithms, which have been used in
recent simulation
studies~\cite{halperin:sigcomm2011,zhou:sigcomm2012,tie:pam2012,zhu2:mobicom2014},
may not be effective; more intelligent algorithms may be required in
complex environments. Further, PHY data rate is not always a good
indicator of higher layer performance.

\begin{figure*}[t]
\centering 
\subfigure[Topology 1] {
\label{fig:top1}
\includegraphics[width=2.0in,height=1.4in,angle=0]{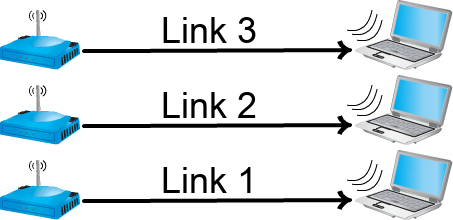} 
}
\subfigure[Topology 2] {
\label{fig:top2}
\includegraphics[width=2.0in,height=1.4in,angle=0]{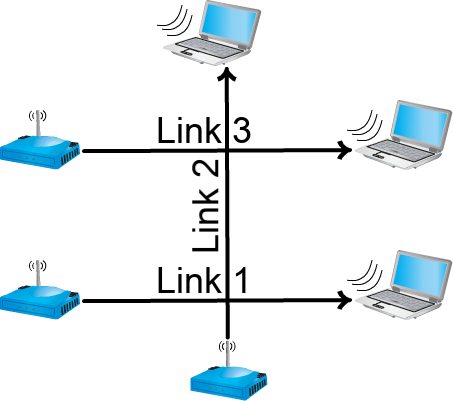}
} 
\subfigure[Topology 3] {
\label{fig:top3}
\includegraphics[width=2.0in,height=1.4in,angle=0]{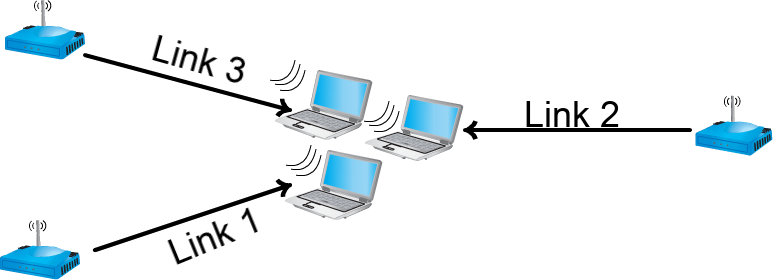} 
}
\vspace{-0.1in}
\caption{Spatial Reuse Topologies}
\label{fig:exp_setup}
\end{figure*}

\section{Spatial Reuse}
\label{sec:reuse}

One of the major advantages of the 60 GHz technology is spatial reuse,
thus enabling efficient use of the channel airtime. Although the use
of directional links makes it harder for sensing-based MAC (CSMA/CA)
to work as desired, it allows for concurrent transmissions on the same
channel. In this section, we use three representative topologies to
look at the degree of spatial re-use allowed by our hardware which
uses wide beams of $30^{\circ}-40^{\circ}$.

Each topology consists of three links inside our lab, which houses 22
cubicles and is full of office-like furniture (chairs, desks, computer
systems etc.). Each of the links is setup at a height of 4 ft and a
LOS path exists between the Tx and Rx. In topology 1
(Figure~\ref{fig:top1}), the links are placed parallel with separation
of 8 ft. between consecutive links. For topology 2 link 2 is
perpendicular to links 1 and link 3. Topology 3 is emulates a case
where multiple users are co-located and are serviced by three
different APs. All three links operate on the same channel. For each topology, we
measure TCP throughput of each individual link when it is operating
alone and when all three links are active at the same time. To
quantify the degree of spatial re-use, we use the spatial reuse factor
$\beta$ introduced in~\cite{sur:sigmetrics2015}, which is equal to the
sum throughput of concurrent links divided by the average throughput
of isolated links.\footnote{$\beta$ has a max. value of 3 (when there
  is no mutual interference). Higher value indicates better spatial
  reuse.} 

Table~\ref{tab:reuse} summarizes the results. Topology 1, where links
are distant enough that side lobes do not cause interference, provides
for maximum spatial reuse. Topology 2 provides the least spatial reuse
as Links 1 and 3 get around 60\% of their isolated throughput. In
topology 3, spatial reuse is better than in topology 2, even though
the receivers of the three links are located very close to each
other. On the other hand, spatial reuse for 802.11ac is close to 1 as it does not
allow for concurrent transmissions.

\begin{table}[h]
\centering
\vspace{-0.2in}
\caption{Spatial Reuse Factor ($\beta$)}
\label{tab:reuse}
\begin{tabular}{l|c|c|c|c|}
\cline{2-5}
                                    & {\bf Topology 1} & {\bf Topology 2} & {\bf Topology 3} & {\bf 802.11ac} \\ \hline
\multicolumn{1}{|c|}{{\bf $\beta$}} & 2.95             & 2.16             & 2.78             & 0.92           \\ \hline
\end{tabular}
\vspace{-0.1in}
\end{table}

\noindent{\bf Remarks} The results demonstrate a large advantage of
802.11ad over legacy WiFi. Even with imperfect hardware using wide
beams and sidelobes, spatial reuse is much higher than in the case of
802.11ac. This result combined with the higher data rates of 802.11ad
compared to 802.11ac and the very small form factors of mmWave
technology make 802.11ad a strong candidate for delivering the
multi-fold throughput increase in future WLANs.

\section{Impact of Human Blockage}
\label{sec:human}

The inability of 60 GHz links to pass through human body without
suffering severe attenuation, owing to the small wavelength of the
carrier wave, is often described as one of the major challenges for
the mmWave technology.  A human in the LOS between the transmitter and
the receiver can attenuate the signal by 20-30
dB~\cite{singh:jsac2009}, resulting in link
outage.~\cite{zhu:mobicom2014} found that in outdoor picocell settings
the impact of static pedestrians is limited in a very small area
around the user due to the base station height (6 m). However, the
impact of groups of moving pedestrians becomes heavier. Recent studies
in indoor settings~\cite{tie:pam2012,sur:sigmetrics2015} showed that
human blockage remains a major challenge, and becomes worse due to the
long re-connection times of existing 802.11ad hardware. We investigate
the severity of this problem through two sets of experiments.
 
\subsection{Controlled experiment} We performed experiments in the
  Hall and Corridor with controlled human placement
  and motion. We study the blockage caused by both permanent human
  obstruction of the LOS link and transient motion which disrupts the
  link temporarily. We are primarily concerned with measuring the
  \textit{reconnection time} and \textit{throughput degradation}.

At each location, we tried four different Tx-Rx distances (8'6'',
16'6'', 24'6'', and 32'6''). At each location, we considered two kinds
of blockage: mobile and static. In the former, a person walks in
random fashion along the LOS path between the Tx and Rx. If the link
breaks as a result of such motion, the human moves away from the link
to allow it to recover. In the latter, a person stands permanently
between the Tx and Rx, forcing the Tx to find an alternate NLOS
path. In cases where the link does not break in spite of the blockage,
we measured the throughput degradation caused due to human
presence. Table~\ref{tab:blockage} summarizes our findings, plotting
the average values over 5 experiments. A zero reconnection time
indicates that the link did not break; in that case, the value in
brackets indicates the observed TCP throughput. An infinity ($\infty$)
value means that the Tx was unable to re-establish the link even
several minutes after introducing the blockage.

\begin{table}[h]{\scriptsize
\centering
\vspace{-0.2in}
\caption{Reconnection time and throughput in case of human blockage (M:mobile, S:Static).}
\label{tab:blockage}
\begin{tabular}{l|c|c|c|c|c|c|c|c|}
\cline{2-9}
                                     & \multicolumn{2}{c|}{{\bf 8'6''}}                                     & \multicolumn{2}{c|}{{\bf 16'6''}}                                                         & \multicolumn{2}{c|}{{\bf 24'6''}}                                                         & \multicolumn{2}{c|}{{\bf 32'6''}}                                     \\ \cline{2-9} 
                                     & {\bf M}                                          & {\bf S} & \multicolumn{1}{l|}{{\bf M}}                     & \multicolumn{1}{l|}{{\bf S}} & \multicolumn{1}{l|}{{\bf M}}                     & \multicolumn{1}{l|}{{\bf S}} & \multicolumn{1}{l|}{{\bf M}} & \multicolumn{1}{l|}{{\bf S}} \\ \hline
\multicolumn{1}{|c|}{{\bf Corridor}} & \begin{tabular}[c]{@{}c@{}}0\\ {[}573{]}\end{tabular} & 0            & \begin{tabular}[c]{@{}c@{}}0\\ {[}728{]}\end{tabular} & $\infty$                          & \begin{tabular}[c]{@{}c@{}}0\\ {[}441{]}\end{tabular} & $\infty$                          & 16.61                             & $\infty$                          \\ \hline
\multicolumn{1}{|l|}{{\bf Hall}}     & 15.34                                                 & $\infty$     & 16.07                                                 & $\infty$                          & 15.42                                                 & $\infty$                          & 16.67                            & $\infty$                          \\ \hline
\end{tabular}}
\vspace{-0.1in}
\end{table}

In the Corridor, the link was resilient to transient human obstruction
for distances upto 24'6''. For a Tx-Rx distance of 8'6'', the link did
not break even in the presence of permanent human blockage. Note
though that throughput, which was always higher than 800 Mbps before
the appearance of the human, dropped significantly in some cases. For
larger distances, the Tx failed to find an alternative path in the
presence of permanent human blockage. Note that in order to simulate
the worst case, the person in most cases stood very close to the Tx or
Rx making it harder for the Tx to search for a new path. On the other
hand, human blockage introduced mid-way between the Tx and Rx failed
to break the link. Lastly, for even greater distance (32'6''), the
link failed under human mobility and it took upto 16.6 seconds for it
to be established again.
 
The results in the Hall are very different. The link broke under human
mobility for all distances and reconnection times were in the range of
15.34 to 15.67 seconds, similar to the Corridor case, and much longer
than the values reported in~\cite{tie:pam2012,zhu:mobicom2014}. The
absence of any reflective surface in the immediate vicinity of the Tx
made it harder for Tx to quickly find an alternative path when it was
blocked momentarily. In case of static blockage, the Tx failed to find
a NLOS path in all cases.

\noindent{\bf Remarks} The results of the controlled experiments
indicate that a static human near the Tx or Rx presents a much bigger
challenge compared to transient blockages introduced by human
motion. Nonetheless, the reconnection times are extremely high when
considered in the context of multi-Gbps throughput. One potential way
to address this problem is beam dilation,
although~\cite{sur:sigmetrics2015} showed that it only works under
high SNR scenarios. Hence, there is a need for faster, more efficient
rebeamforming algorithms, potentially combined with mechanisms that
distinguish the cause of link outage (human blockage or client
mobility), as different approaches work better in each
scenario~\cite{sur:sigmetrics2015}.

\subsection{In the wild experiment} 

\begin{figure*}[]
\centering 
\subfigure[1 AP] {
\label{fig:cdf_1}
\includegraphics[width=2.2in,height=1.6in,angle=0]{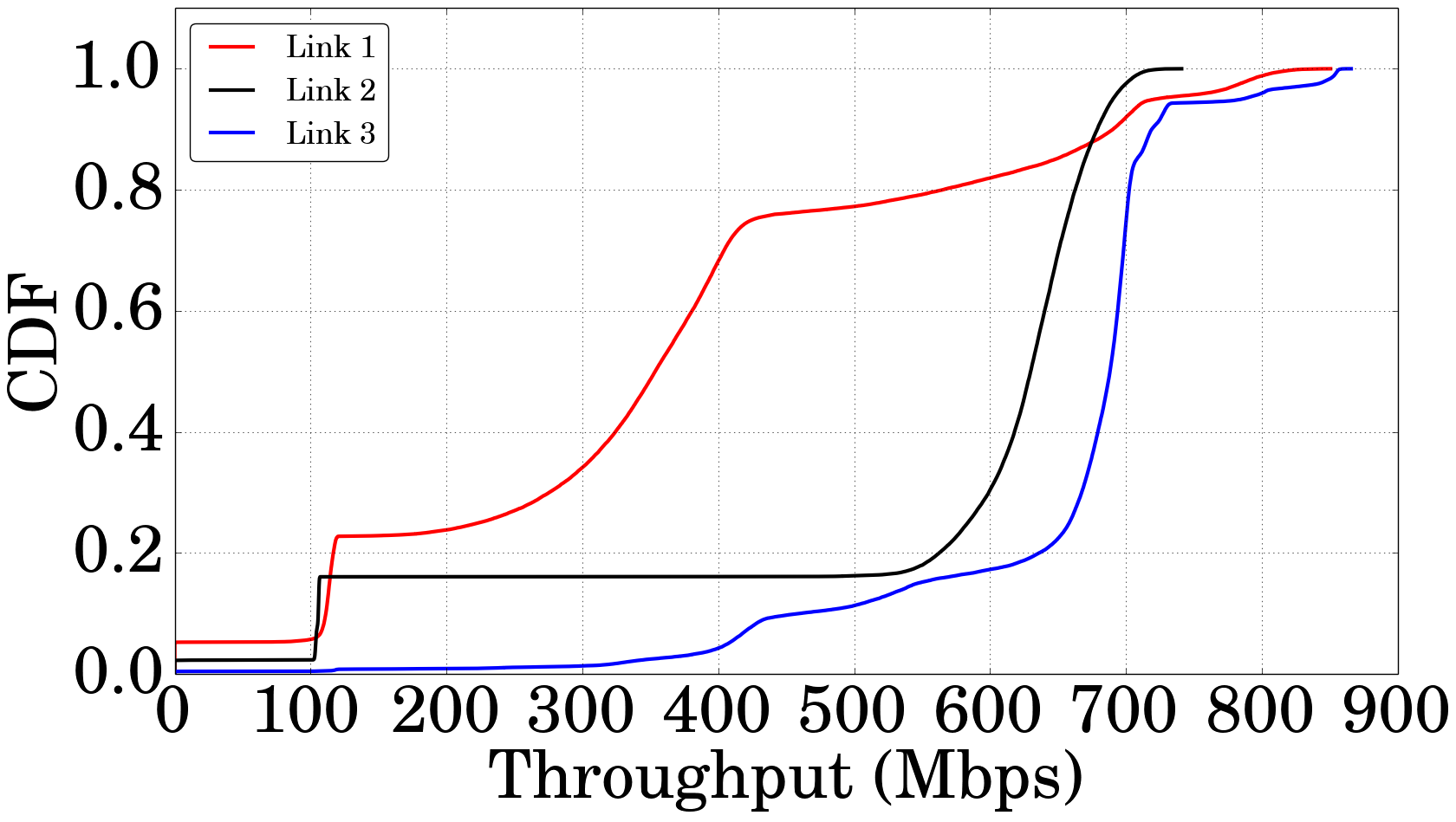} 
}
\subfigure[2 APs] {
\label{fig:cdf_2}
\includegraphics[width=2.2in,height=1.6in,angle=0]{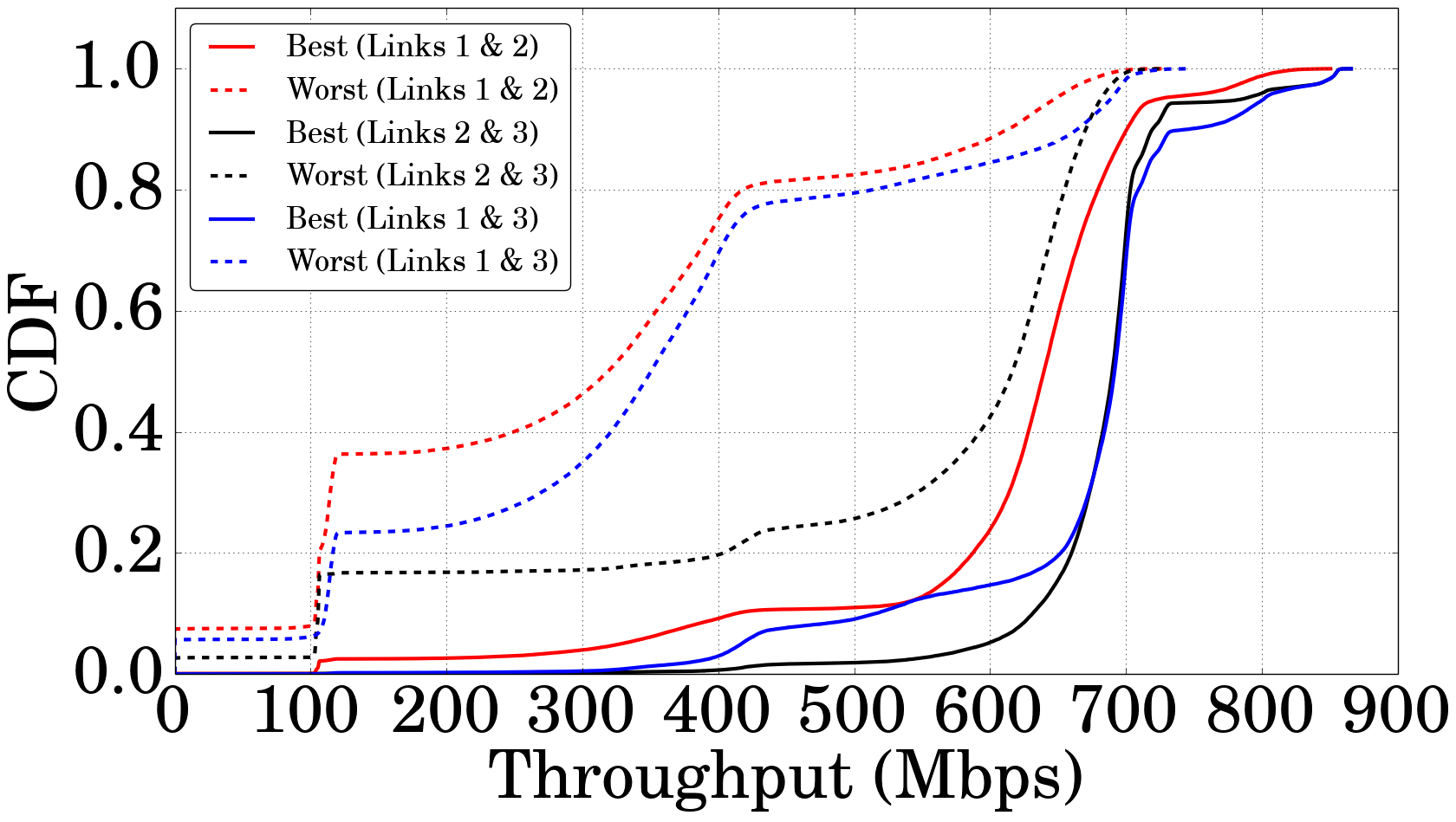}
} 
\subfigure[3 APs] {
\label{fig:cdf_3}
\includegraphics[width=2.2in,height=1.6in,angle=0]{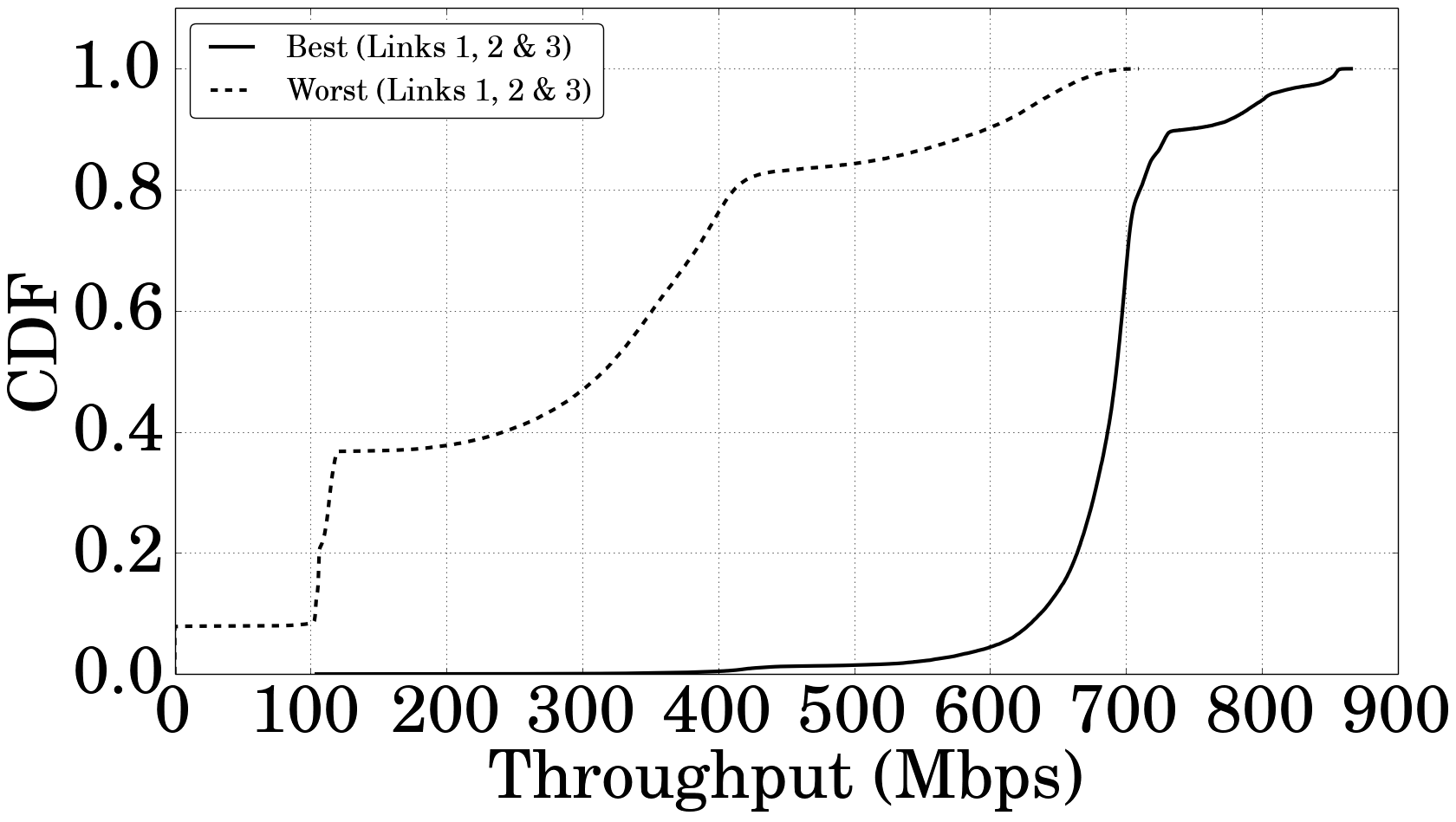} 
}
\caption{CDF of TCP throughput over 15 hours.}
\label{fig:exp_setup}
\end{figure*}

The controlled experiments gave an insight into the challenges arising
out of the presence of humans into the environment.  However, two
questions remain unanswered: (i) how often would such blockages occur
in a typical WLAN and (ii) can serving a client with multiple APs
(similar to the BS picocloud scenario in~\cite{zhu:mobicom2014} for 60
GHz outdoor picocells) help mitigate this problem? To answer these
questions, we used a methodology similar to that
of~\cite{zhu:mobicom2014} since Wilocity radios do not allow switching
between APs on-the-fly. We deployed three links in our lab, very close
to each other, emulating a single client which can potentially connect
to any of the three docks/APs, in a topology very similar to that of
Topology 3 (Figure~\ref{fig:top3}) for a period of 15 hours. The 15
hour experiment period included both night hours and day hours of the
following day. We recorded per second TCP throughput for each of the
three links.

Figures \ref{fig:cdf_1}, \ref{fig:cdf_2}, and \ref{fig:cdf_3} present
the CDF of throughput in three cases when one, two, or three APs are
considered to be deployed. In the first case, where we assume that
only one of the 3 APs was available for connection to the client, we
see that each of the links was blocked/disconnected (zero throughput)
for less than 5\% of the time and two of the links maintained a
throughput between 600 and 700 Mbps most (around 70\%) of the
time. However, throughput above 800 Mbps was achieved for less than
5\% of time by each link.


When considering 2 APs, we have 3 possible combination of
APs. Further, for each combination, we plot both the best throughput
achieved out of the two APs and the worst one for
comparison. Interestingly, all the three combinations gave a 0
percentage of disconnection time, when considering best throughput
scenario, indicating that two APs would have been sufficient for
maintaining 100\% uptime. 


For the 3-AP case, we show the best and worst throughput CDFs. If a
client were to connect to the best AP all the time, it would never
experience disconnection and would maintain a median throughput of
around 680 Mbps.



\noindent{\bf Remarks} Our 15 hour experiment showed that the presence
of humans in a typical office environment does not have a significant
impact on connectivity but can cause a significant throughput
degradation. However, using 2 APs per client can provide 100\% uptime
and high throughput. 


\section{Conclusions}
In this paper, we investigated the feasibility of building
general-purpose indoor multi-gigabit 60 GHz WLANs by conducting an
extensive measurement study in a typical academic office building
using commercial 802.11ad hardware. We studied the range of 60 GHz
transmissions in indoor environments, and the impact of antenna
height, location, orientation, and distance on 60 GHz performance at
multiple layers of the protocol stack. We explored the interaction
between link layer indicators and transport layer throughput. We also
studied spatial reuse opportunities and the impact of human
blockage. Our results indicate that the 60 GHz technology can indeed
be a viable option for building multi-gigabit WLANs but there are
several challenges that need to be addressed before 60 GHz WLANs can
be realized in practice.

%
\bibliographystyle{abbrv}
\bibliography{wireless}

%

\end{document}